\documentclass[12pt]{iopart}
\pdfoutput=1
\usepackage[utf8]{inputenc} 			
\usepackage[T1]{fontenc}				
\usepackage{lmodern}					
\usepackage[pdftex]{graphicx} 		    
\usepackage[svgnames]{xcolor}
\usepackage{cite}
\usepackage{iopams}                     
\usepackage{setstack}                   
\usepackage[english]{babel}
\usepackage{lmodern}
\usepackage{ulem}

\usepackage{listings}
\usepackage{tikz}
\graphicspath{{./}{./Images/}}
\newcommand{\eqref}[1]{\eref{#1}}

\definecolor{linkcolor}{rgb}{0,0,0.6}

\newcommand\redsout{\bgroup\markoverwith{\textcolor{red}{\rule[0.5ex]{2pt}{0.4pt}}}\ULon}

\usepackage[percent]{overpic}

\usepackage[pdftex,colorlinks=true,pdfstartview=FitV,linkcolor= linkcolor,citecolor= linkcolor,urlcolor= linkcolor,hyperindex=true,hyperfigures=false]{hyperref}
\begin{document}

\title[A run-and-tumble particle around a spherical obstacle]{A run-and-tumble particle around a spherical obstacle: steady-state distribution far from equilibrium}
\author{Thibaut Arnoulx de Pirey$^1$}
\author{Fr\'ed\'eric van Wijland$^2$}
\address{$^1$ Department of Physics, 
			Technion-Israel Institute of Technology,
		    Haifa 32000, 
		    Israel}
\address{$^2$ Laboratoire Matière et Systèmes Complexes,
              Université Paris Cité \& CNRS (UMR 7057),
              10 rue Alice Domon et Léonie Duquet,
              75013 Paris,
              France}
\ead{t.depirey@campus.technion.ac.il}
\date{\today\ -- \jobname}

\begin{abstract}
We study the steady-state distribution function of a run-and-tumble particle evolving around a repulsive hard spherical obstacle. We show that the well-documented activity-induced attraction translates into a delta peak accumulation at the surface of the obstacle accompanied with an algebraic divergence of the density profile close to the obstacle. We obtain the full form of the distribution function in the regime where the typical distance run by the particle between two consecutive tumbles is much larger than the size of the obstacle. This provides an expression for the low-density pair distribution function of a fluid of highly persistent hard-core run-and-tumble particle. This also provides an expression for the steady-state probability distribution of highly-ballistic active Brownian particles and active Ornstein-Ulhenbeck particles around hard spherical obstacles.   \\

\end{abstract}
\submitto{\JSTAT}

\section{Introduction}

Such spectacular collective physical phenomena as collective alignment~\cite{PhysRevLett.75.1226,gregoire2004onset,ramaswamy-2010} or motility-induced phase separation~\cite{PhysRevLett.100.218103,redner2013structure,PhysRevLett.108.235702,cates2015motility} are trademarks of interacting active, self-propelled, particles. However, the individual behavior of an active particle interacting with an external potential displays equally puzzling features. To name but a few: a steeper decay of density for sedimenting run-and-tumble particles (RTP) under a constant gravity field~\cite{Tailleur_2009}, a depleted probability at the bottom of a harmonic well for RTPs~\cite{Tailleur_2009, pototsky2012active,  solon2015active, fodor2018non} or active Brownian particles (ABP),
and more prominently perhaps, a tendency to be attracted to otherwise repulsive obstacles. This phenomenon was for instance explored in two-dimensional channels for RTPs~\cite{ezhilan2015distribution} or ABPs~\cite{wagner2017steady}, and more generally for particles confined within walls~\cite{lee2013active,elgeti2015run}. Physically more complex systems have been considered, such as a needle-like particle interacting with a spherical obstacle~\cite{spagnolie2015geometric} in the presence of a surrounding fluid, which adds hydrodynamic interactions to an already nontrivial problem. The understanding of the one-particle problem is also instrumental in breaking down the mechanisms at work in many-body collective phenomena into elementary ones. In systems of particles interacting via pairwise forces considered in the dilute limit one can, as in the virial approach familiar in equilibrium systems, focus on a single pair of particles. For that pair of particles, the associated reduced particle~\cite{goldstein1950classical} evolves in an effective external potential. However, the statistics of the active self-propulsion force felt by the reduced particle need not --and will in general not-- be identical to that exerted on each individual particle. 

From a purely analytical standpoint, instances in which the stationary distribution can be found exactly are scarce. A single RTP in one space dimension, in an arbitrary external potential or with spatially varying tumble rate and speed~\cite{van1984activation,schnitzer1993theory,solon2015Notpressure,dhar2019run,thompson2011lattice} is a notable exception. This problem was also solved on a one-dimensional lattice and in the continuum for a RTP in between two hard walls~\cite{PhysRevLett.116.218101,malakar2018steady}. For the more realistic case of two hard-disk RTPs, an approximate treatment was recently put forward~\cite{speck-kirkwood}. The goal of the present work is to address the case of a RTP in contact with a hard spherical obstacle in space dimension $d$ with a particular emphasis on the highly ballistic limit where the persistence length $\ell_p$  of the RTP is much larger than the obstacle size $\sigma$. The persistence length, which reads $\ell_p = v_0 \tau$ with $\tau$ the correlation of the self-propulsion velocity and $v_0$ its amplitude, is the mean distance run by the free particle between two consecutive tumbles. In the highly-ballistic regime, the particle thus almost never tumbles when in the vicinity of the obstacle. Incidentally, the highly ballistic limit turns out to be relevant for studying collective phenomena such as the motility-induced phase separation~\cite{PhysRevLett.108.235702}. This regime has also been studied for its connections with sheared granular systems~\cite{Agoritsas21jsm,Mo20sm}.

Our main results are as follows. For all $\eta = \sigma/\ell_p$, the stationary distribution function is shown to exhibit a delta peak accumulation at contact and a density profile away from the obstacle (in the bulk of the system) that diverges at $r = \sigma$. The structure of this divergence is elucidated. We furthermore prove that the bulk distribution function in the position and self-propulsion space exhibits a delta peak accumulation along typical trajectories of the stochastic dynamics that we characterize. The steady-state distribution is then fully characterized in $d \geq 2$ as $\eta \to 0$. Corrections to this limiting behavior to next order in $\eta$ are also investigated and we show that the amplitude of the delta peak accumulation decreases as the persistence length is reduced. We lastly show that in the highly ballistic limit $\eta \to 0$, far-field hydrodynamic interactions, as captured by the Oseen tensor, reduce the amplitude of the delta peak accumulation.

We begin, in Sec.~\ref{sec:Model}, by introducing the particle model we consider and show obtain coupled equations for the bulk and surface contributions to the steady-state distribution. In Sec.~\ref{subsec:Sol}, we then show that, in steady-state, the marginal density in position space is the solution to an integral equation. This equation is solved in the vicinity of the obstacle in Sec.~\ref{sec:vicinity} and in the highly ballistic regime in Sec.~\ref{sec:ballistic}.  The effect of hydrodynamic interactions in the highly ballistic regime is investigated in Sec.~\ref{sec:Hydro}. Our conclusion in Sec.~\ref{sec:conclusion} gathers possible paths deserving further exploration.

\section{The steady-state distribution: bulk and surface contributions}\label{sec:Model}

In this section, we derive the equations satisfied by the steady-state probability distribution of a run-and-tumble particle evolving in the presence of a hard spherical obstacle in arbitrary space dimension $d>1$. We show that the latter splits into a bulk contribution and a surface contribution localized at the surface of the obstacle. We obtain coupled equations relating these two contributions. Later sections are devoted to reformulating these equations in terms of an integral equation for the density field and which we solve in some limiting cases.

To achieve this, we start with the equation of motion of an RTP evolving in a smooth spherically symmetric repulsive potential. The hard-sphere limit is carefully taken at a later stage. The position of the RTP evolves according to
\begin{eqnarray}
\frac{\rmd \bi{r}}{\rmd t} = v_0 \bi{u}(t) - \bnabla V(\bi{r}) \, ,
\label{eq:EOM}
\end{eqnarray}
where $\bi{u}(t)$ is a unit vector that is uniformly reoriented on the unit sphere with rate $\tau^{-1}$. The potential is assumed to depend only on the distance $r = ||\bi{r}||$ so that $V(\bi{r}) = V(r)$. A cartoon depicting the interaction between an incoming particle and a hard obstacle is shown in Fig.~\ref{fig:collision}. Note that in Fig.~\ref{fig:collision}, both the particle and the obstacle have the same size. In practice, the results only depend on the exclusion radius of the hard-sphere potential, meaning the radius of the obstacle plus the radius of the particle. Figure \ref{fig:collision} further illustrates the key difference with the physics of one-dimensional systems (or higher-dimensional ones in the presence of an extended flat wall). Here, the particle can leave the vicinity of the obstacle even in the absence of tumbling event. The stationary state distribution function $P(\bi{r},\bi{u})$ solves the master equation,
\begin{figure}
\begin{center}
\begin{overpic}[totalheight=4.5cm]{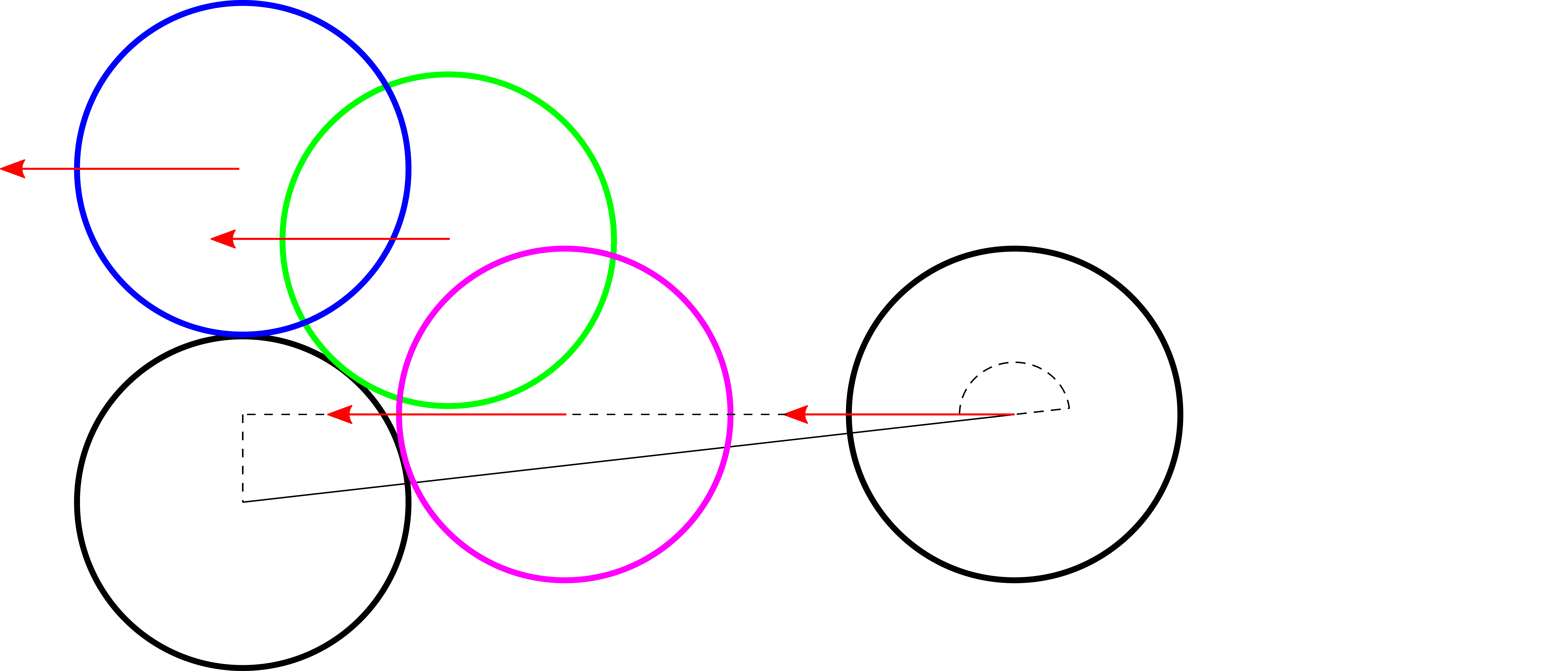}

 \put (13,12.5) {\textcolor{black}{$b$}}
 \put (57,17) {\textcolor{red}{$\bi{u}$}}
 \put (2,33) {\textcolor{red}{$\bi{u}$}}
 \put (64,21) {$\theta$}
 \put (49,12){$r$}
\end{overpic}
  \caption{A collision of an active hard-sphere (black, rightmost) with diameter $\sigma$ and impact parameter $b=r\sin\theta<\sigma$ (with $\cos\theta <0$) onto a pinned (black, leftmost) one. In the following we introduce the notation $w = \cos\theta$. The incoming particle with direction $\bi{u}$ hits the target sphere (at the magenta position) and then skids around, as in the green position. In the absence of tumbling, it eventually takes off at the blue position when its orientation $\bi{u}$ becomes tangent to the target sphere. In the highly ballistic limit, no tumble can occur over the typical skidding distances.}\label{fig:collision}
\end{center}
\end{figure}
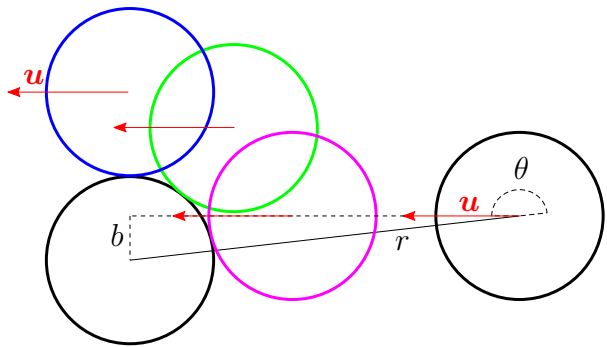
\begin{eqnarray}\label{eq:FP_1part_1obs_vect}
\fl -v_0\bi{u} \cdot\bnabla_{\bi{r}}P(\bi{r},\bi{u}) + \bnabla_{\bi{r}}\left(P(\bi{r},\bi{u})\bnabla_{\bi{r}}V(\bi{r})\right) + \frac{1}{\tau}\left(\int \frac{\rmd \bi{u}'}{\Omega_d}P(\bi{r},\bi{u}') - P(\bi{r},\bi{u})\right) = 0 \, .
\end{eqnarray}
We take advantage of the rotational symmetry to write $P(\bi{r},\bi{u}) = P(r,w)$ where $w = \bi{r} \cdot \bi{u} / r = \cos\theta$ with $\theta$  the polar angle measured with respect to $\bi{u}$. After writing Eq.~\eqref{eq:FP_1part_1obs_vect} using these spherical coordinates, we obtain
\begin{eqnarray} \fl
   0 =  -v_0 \left( w \partial_r P(r,w) + \frac{1-w^2}{r}\partial_w P(r,w)\right) + \frac{1}{r^{d-1}}\partial_r\left(r^{d-1} P(r,w) \partial_r V(r) \right) \nonumber \\  + \frac{1}{\tau}\left(\int_{-1}^{1} \frac{\rmd w'}{2 W_{d-2}}\left(1-w'^2\right)^{\frac{d-3}{2}} P(r,w') - P(r,w)\right) \, .
\label{eq:FP}
\end{eqnarray}
Here, the normalization constant $W_n$ is the $n^{\rm{th}}$ Wallis integral,
\begin{equation}
W_n = \frac{1}{2}\int_{-1}^{1} \rmd w' \left(1-w'^2\right)^{\frac{n-1}{2}} \,.
\end{equation}
The equations of motion written in terms of the variables $r,\,w$ are discussed in \ref{app:dynvar}.
One of the ways of ultimately implementing a hard-sphere potential with  exclusion diameter $\sigma$ is to begin with a smoothly repulsive potential $V(r)$ of the form
\begin{eqnarray}    
        V(r) = V_0 \, \rme^{-\frac{r-\sigma}{\epsilon\sigma}} \, ,
\end{eqnarray}
and to take the hard sphere limit $\epsilon \to 0^+$ by following the program explained in \cite{de2019active} and further detailed in what follows. For any fixed $r > \sigma$, we first define the bulk distribution function $f(r,w)$ by
\begin{equation}
f(r,w) = \lim_{\epsilon \to 0^+} P(r,w) \,.
\end{equation}
It follows from Eq.~\eqref{eq:FP} that
\begin{eqnarray} \fl
   0 =  -v_0 \left(w \partial_r f(r,w) + \frac{1-w^2}{r}\partial_w f(r,w)\right) \nonumber \\  + \frac{1}{\tau}\left(\int_{-1}^{1} \frac{\rmd w'}{2 W_{d-2}}\left(1-w'^2\right)^{\frac{d-3}{2}} f(r,w') - f(r,w)\right) \, ,
\label{eq:FP_bulk}
\end{eqnarray}
which expresses the flux balance between the free streaming of the particle for $r > \sigma$ under the action of the active self-propulsion and the random reorientation of the latter. We now turn to the behavior of the distribution function close to $r=\sigma$ and show that it develops a delta singularity when $\epsilon \to 0^+$. From the equation of motion Eq.~\eqref{eq:EOM}, we know that $\forall \epsilon > 0$, $P(r,w) = 0$ for $r \leq r^*$ with $r^*$ defined by $V'(r^*) = - v_0$. Here $r^* - \sigma = \epsilon \sigma \ln\left(\epsilon v_0/V_0\right)$ goes to 0 as $\epsilon \to 0^+$. To identify the singularity, we define the surface distribution function as
\begin{equation}
\Gamma(w) = \sigma^{-1} \lim_{r\to\sigma}\lim_{\epsilon \to 0^+} \int_{r^*}^r \rmd r' P(r',w) \,.
\end{equation}
We then multiply Eq.~\eqref{eq:FP} by $r^{d-1}$ and integrate it between $r^*$ and some fixed $r > r^*$. After integration by part and using the boundary condition $P(r^*,w) = 0$, we obtain
\begin{eqnarray} \fl
    0 =  -v_0 w r^{d-1}P(r,w) +  v_0 (d-1) w \int_{r^*}^r \rmd r' r'^{d-2} P(r',w) \nonumber \\ \fl \phantom{0=} + r^{d-1} P(r,w)V'(r)   - v_0 (1-w^2) \partial_w \int_{r^*}^r \rmd r' r'^{d-2} P(r',w)  \nonumber \\ \fl \phantom{0=} + \frac{1}{\tau}\int_{r^*}^r \rmd r' r'^{d-1}\left(\int_{-1}^{1} \frac{\rmd w'}{2 W_{d-2}}\left(1-w'^2\right)^{\frac{d-3}{2}} P(r',w') - P(r',w)\right) \, ,
\label{eq:FPepsbis}
\end{eqnarray}
At fixed $r$, we take the limit $\epsilon \to 0^+$ in which the term proportional to $V'(r)$ vanishes. We then take the $r \to \sigma$ limit, yielding
\begin{eqnarray} \fl
    0 =  - w f(\sigma,w) +  (d-1) w \Gamma(w)  - (1-w^2)\Gamma'(w) \nonumber \\ \fl \phantom{0=}  + \frac{\sigma}{v_0\tau}\left(\int_{-1}^{1} \frac{\rmd w'}{2 W_{d-2}}\left(1-w'^2\right)^{\frac{d-3}{2}} \Gamma(w') - \Gamma(w)\right) \, ,
\label{eq:FPsurf1}
\end{eqnarray}
The flux of particles with radial self-propulsion $w$ arriving on the surface of the obstacle from the bulk is proportional to $w f(\sigma,w)$ and is accounted for in the first term of Eq.~\eqref{eq:FPsurf1}. A non-zero flux from the bulk implies a non-vanishing surface distribution function $\Gamma(w)$ as seen in Eq.~\eqref{eq:FPsurf1}. It is clear from Eq.~\eqref{eq:FP_bulk} that $f(\sigma, w)$ cannot be zero for all $w$ since such a boundary condition would imply a vanishing bulk distribution for all $r > \sigma$. Thus the probability distribution function develops in the hard sphere limit a singular part at contact in the form of a delta peak at $r = \sigma$.

To summarize, denoting $z = r/\sigma$, the stationary distribution function takes the following form
\begin{eqnarray}
P(\bi{r},\bi{u}) = f(z,w) +  \Gamma(w) \delta (z-1) \, .
\end{eqnarray}
We have obtained two coupled integro-differential equations satisfied by $f(z,w)$ and $\Gamma(w)$. They depend on a single dimensionless parameter $\eta = \frac{\sigma}{\ell_p}$ comparing the obstacle size to the persistence length $\ell_p=v_0\tau$. First, the bulk distribution function obeys for $z>1$,
\begin{eqnarray} \label{eq:FPbulk}
 w \partial_z f + \frac{1-w^2}{z}\partial_w f + \eta f = \eta \rho(z) \, .
\end{eqnarray}
with
\begin{eqnarray}\label{eq:def_rho}
\rho(z) = \int_{-1}^{1} \frac{\rmd w}{2 W_{d-2}} (1-w^2)^{\frac{d-3}{2}}f(z,w) \, ,
\end{eqnarray}
the bulk density. Second, the surface distribution function follows from
\begin{eqnarray}\fl
\Gamma'(w) - \frac{w}{1-w^2}(d-1) \Gamma(w) + \frac{\eta}{1-w^2}\Gamma(w) = - \frac{w}{1-w^2}f(1,w) + \frac{\eta}{1-w^2}\hat{\Gamma} \, ,
\label{eq:FPsurf}
\end{eqnarray}
with 
\begin{eqnarray} 
\hat{\Gamma} = \int_{-1}^{1} \frac{\rmd w}{2 W_{d-2}} (1-w^2)^{\frac{d-3}{2}}\Gamma(w) \, ,
\end{eqnarray}
the surface density. We lastly show that the flux balance equation at the surface of the obstacle in Eq.~\eqref{eq:FPsurf} implicitly contains a boundary condition for the bulk distribution function. We can indeed show that $\Gamma(w > 0) = 0$ which expresses the fact that particles whose self-propulsion points away from the obstacle do not accumulate at its surface. Hence, Eq.~\eqref{eq:FPsurf} thus yields the boundary condition
\begin{equation} \label{eq:surfaceBC}
f(1, w > 0) = \frac{\eta \hat{\Gamma}}{w} \, .
\end{equation}
As mentioned before, the flux of particles going from the surface of the obstacle to the bulk with orientation $w > 0$ is proportional to $w f(1, w)$. Equation \eqref{eq:surfaceBC} shows that this flux is independent of the orientation $w$. This is expected since this flux is uniquely generated by isotropic tumbling events. Indeed, in the absence of tumbling, the particle leaves the obstacle only when $w = 0$, see Fig.~\ref{fig:collision}.

To show that indeed $\Gamma(w > 0) = 0$, we start by regularizing the product $P(r,w)V'(r)$ in the hard sphere limit. For that, we integrate once more Eq.~\eqref{eq:FPepsbis} over $r \in [r^*,r']$ thus yielding,
\begin{eqnarray} \fl
   0 =  -v_0 w  \int_{r^*}^{r'} \rmd r \, r^{d-1}P(r,w) +   \int_{r^*}^{r'} \rmd r \, r^{d-1} P(r,w)V'(r) \nonumber \\ \fl \phantom{0=} + \int_{r^*}^{r'} \rmd r \int_{r^*}^r \rmd r'' \, r''^{d-2} \,\Bigg[ v_0 (d-1) w P(r'',w) - v_0 (1-w^2) \partial_w P(r'',w)  \nonumber \\ \fl \phantom{0=} + \frac{r}{\tau}\left(\int_{-1}^{1} \frac{\rmd w'}{2 W_{d-2}}\left(1-w'^2\right)^{\frac{d-3}{2}} P(r'',w') - P(r'',w)\right)\Bigg]  \, .
\label{eq:FP_regul_v}
\end{eqnarray}
We then take the $\epsilon \to 0^+$ limit at fixed $r'$ and afterwards we take the $r' \to \sigma$ limit. The double integral term vanishes and we are left with 
\begin{eqnarray}
\label{eq:regul_V}\fl
\lim_{r' \to \sigma} \lim_{\epsilon \to 0^+} \int_{r^*}^{r'} \rmd r P(r,w) V'(r)  = v_0 w  \lim_{r \to \sigma} \lim_{\epsilon \to 0^+} \int_{r^*}^{r'} \rmd r P(r,w)   = v_0 \sigma w \Gamma(w)
\end{eqnarray}
proving that $\Gamma(w > 0) = 0$ as $V'(r)<0$ for all $r$.

\section{An integral equation over the density field}\label{subsec:Sol}

We have derived in the hard-sphere limit the two coupled equations satisfied by the bulk and the surface distribution functions, respectively $f(z,w)$ and $\Gamma(w)$. In this section, we take advantage of the fact that Eq.~\eqref{eq:FPbulk} satisfied by $f(z,w)$ is first order in the partial derivatives $\partial_z$ and $\partial_w$ and use the methods of characteristics to recast it into an integral equation for the spatial density field $\rho(z)$ only. We furthermore prove that the bulk distribution function $f(z,w)$ exhibits a delta peak accumulation along typical trajectories of the stochastic dynamics that we characterize.

\begin{figure}
\begin{center}
\begin{overpic}[width=.8\columnwidth]{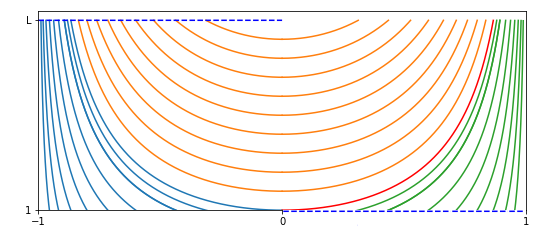}

 \put (51.4, 2) {\textcolor{black}{$w$}}
 \put (3,25) {\textcolor{black}{$z$}}
\end{overpic}
  \caption{Parametric representation of the characteristics of Eq.~\eqref{eq:FPbulk} satisfied by the bulk distribution function. The dashed blue lines correspond to regions of the plane where boundary conditions are specified, see Eq.~\eqref{eq:surfaceBC} and Eq.~\eqref{eq:boundL}. We group the characteristics depending on which boundary condition they intersect and on whether or not they intersect the line $z=1$ representing the surface of the obstacle. (\textbf{Green}) Characteristics that are connected to the boundary condition at $z = 1$. They correspond to domain 1 of the main text. (\textbf{Blue}) Characteristics that are connected to the boundary condition at radial position $z = L$ and intersect the line at $z=1$. (\textbf{Orange}) Characteristics that are connected to the boundary condition at radial position $z = L$ and do not intersect the line at $z=1$. In the main text, domain 2 is comprised of all the characteristics restricted to $w<0$ (both blue and orange) and domain 3 is comprised of the orange characteristics restricted to $w>0$. (\textbf{Red}) The only characteristic that originates from $\left(z = 1, w = 0\right)$ restricted to $w>0$. After each collision, the particle leaves the obstacle exactly along this line (except when it has flipped in course of skidding), see Fig.~\ref{fig:collision}. We show that the bulk distribution function develops a singularity along this line which correspond to domain 4 of the main text.}
  \label{fig:karact}
\end{center}
\vspace{-0.10cm}
\end{figure}

The characteristics of Eq.~\eqref{eq:FPbulk} are parametric lines $\left(z(s), w(s)\right)$ such that for any function $g(z,w)$ we have
\begin{equation}
\frac{\rmd g\left(z(s),w(s)\right)}{\rmd s} = w(s) \partial_z g\left(z(s),w(s)\right) + \frac{1-w(s)^2}{z(s)}\partial_w g\left(z(s),w(s)\right) \,.
\end{equation}
They obey
\begin{equation}
\label{cases}
\cases{z'(s) = w(s) \,,\\
w'(s) = \frac{1-w^2(s)}{z(s)} \,,\\}
\end{equation}
and are thus lines such that $z \sqrt{1-w^2} = b = \rm{cst}$ with $b$ the impact parameter (as indicated in Fig.~\ref{fig:collision}). They correspond to purely ballistic trajectories and are parametrically depicted in Fig.~\ref{fig:karact} as the fictitious time $s$ is increased. Equation~\eqref{eq:FPbulk} is furthermore supplemented with the boundary condition 
\begin{equation}\label{eq:boundL}
f(L, w < 0) = 1 \,,
\end{equation}
that implements an homogeneous reservoir of incoming particles at distance $L$ from the obstacle, and where $L$ is sent to infinity at the end of the calculation. As $L \to \infty$, the actual form of the boundary condition at $z = L$ is irrelevant.

We now analyze the behavior of the bulk distribution $f(z,w)$ by dividing the $(w,z)$ plane into four domains as explained in Fig.~\ref{fig:karact}.
\subsection{Domain 1: $w>0$ and $z\sqrt{1-w^2} < 1$}
By definition, along a characteristic, the bulk equation reads
\begin{eqnarray}
f'(s) + \eta f(s) = \eta K(s) \,\, {\rm with} \,\, K(s) = \rho(z(s)) \, ,
\end{eqnarray}
and hence it can be integrated into
\begin{eqnarray}
f(s,b)  = f(s=0,b) \rme^{-\eta s} + \eta \rme^{-\eta s} \int_0^s \rmd s' K(s') \rme^{\eta s'} \, .
\end{eqnarray}
A boundary condition must then be implemented to express $f(s=0,b)$ and the $\left(s,b\right)$ variables that parametrize the characteristic must be replaced with their $\left(z,w\right)$ counterparts. We first solve the equation on the domain $w > 0$ and $z\sqrt{1-w^2} < 1$. This domain is generated by characteristics corresponding to trajectories leaving the obstacle after a collision. These are depicted in green in Fig.~\ref{fig:karact}. Along each characteristic we have
\begin{eqnarray}
 z \sqrt{1-w^2} = b < 1 \, , \nonumber  \\
 \Rightarrow \,\,\,   w(s) = \sqrt{1-\frac{b^2}{z(s)^2}} \, .
\end{eqnarray}
Thus
\begin{eqnarray}
 z'(s) = \sqrt{1-\frac{b^2}{z(s)^2}} \, , \nonumber \\
\Rightarrow \,\,\,  z(s)  \sqrt{1-\frac{b^2}{z(s)^2}} = s + \sqrt{1-b^2} \, , 
\end{eqnarray}
as we choose to parametrize the characteristics such that $z(s=0) = 1$. This leads to
\begin{equation}
\label{eq:var1}
\cases{ z(s,b) = \sqrt{s^2 + 2s\sqrt{1-b^2} + 1} \, , \\
 w(s,b) = \frac{s + \sqrt{1-b^2}}{s^2 + 2s\sqrt{1-b^2} + 1} \, , }
\end{equation}
which can be inverted so as to get
\begin{equation}
\cases{
b  = z\sqrt{1-w^2} \, , \\ 
s  = zw - \sqrt{1-z^2(1-w^2)} \, .}
\end{equation}
Lastly, from Eq.~\eqref{eq:surfaceBC} and Eq.~\eqref{eq:var1}, the boundary condition becomes
\begin{eqnarray}
    f(s = 0, b) = \frac{\eta\hat{\Gamma}}{\sqrt{1-b^2}} \, .
\end{eqnarray}
This finally leads to 
\begin{eqnarray}\label{eq:fdomain1}\fl
f(z,w) =  \, \frac{\eta \hat{\Gamma}}{\sqrt{1-z^2(1-w^2)}}\rme^{-\eta (zw - \sqrt{1-z^2(1-w^2)})} \nonumber \\  \,+  \eta \rme^{-\eta z w}\int_1^z \rmd z' \rho(z') \frac{z'}{\sqrt{z'^2-z^2(1-w^2)}}\rme^{\eta \sqrt{z'^2-z^2(1-w^2)}} \, .
\end{eqnarray}

\subsection{Domain 2: $w<0$}
This domain is generated by all characteristics that are connected to the boundary condition at $z = L$ restricted to $w<0$. These are depicted in blue and orange in Fig.~\ref{fig:karact}. Following the same route as above, we obtain along a characteristic
\begin{equation} \label{eq:varbis2}
\cases{
 z(s,b) = \sqrt{b^2 + \left(s-L\sqrt{1-\frac{b^2}{L^2}}\right)^2} \, , \\
 w(s,b) = \frac{-L\sqrt{1-\frac{b^2}{L^2}}+s}{\sqrt{b^2 + (s-L\sqrt{1-\frac{b^2}{L^2}})^2}} \, ,}
\end{equation}
with $z(s=0,b) = L$. These equations can be inverted to yield
\begin{equation}\cases{
 b = z\sqrt{1-w^2} \, , \\
 s = zw + L\sqrt{1-\frac{z^2(1-w^2)}{L^2}} \, .}
\end{equation}
Hence,
\begin{eqnarray}\fl 
f(z,w)  = \eta \rme^{-\eta z w} \int_z^L \rmd z' \rho(z') \exp{\left\{-\eta z' \sqrt{1- \frac{z^2(1-w^2)}{z'^2}}\right\}}\frac{1}{\sqrt{1- \frac{z^2(1-w^2)}{z'^2}}} \nonumber \\  \qquad  + \rme^{-\eta\left(zw + L\sqrt{1-\frac{z^2(1-w^2)}{L^2}}\right)}\, , \nonumber \\ \fl
\phantom{f(z,w)} \underset{L \to \infty}{=} \eta \rme^{-\eta z w} \int_z^{+\infty} \rmd z' \rho(z') \exp{\left\{-\eta \sqrt{z'^2- z^2(1-w^2)}\right\}}\frac{z'}{\sqrt{z'^2- z^2(1-w^2)}} \, .
\label{eq:fdomain2}
\end{eqnarray}
The above equation shows, as expected, the irrelevance of the boundary condition at $z = L$ as $L$ is sent to infinity.

\subsection{Domain 3: $w>0$ and $z\sqrt{1-w^2} > 1$}
This domain is generated by the continuation of the orange characteristics of Fig.~\ref{fig:karact} studied in the previous paragraph into the $w >0$ half-plane. In this region of the plane we use the continuity requirement at $w=0$ inferred from Eq.~\eqref{eq:fdomain2} as an initial condition at $s=0$. The characteristics are parametrized by
\begin{equation}\cases{
w(s,b)  = \frac{s}{\sqrt{b^2 + s^2}} \, , \\
z(s,b)  = \sqrt{b^2 + s^2} \, , }
\end{equation}
as $w(s=0,b) = 0$. These equations can be inverted to yield
\begin{equation}\cases{
 b = z\sqrt{1-w^2} \, , \\
 s = zw \, .}
\end{equation}
Furthermore, continuity imposes that 
\begin{eqnarray}
f(s=0,b) = \eta \int_b^{+\infty} \rmd z' \rho(z') \exp{\left\{-\eta z' \sqrt{1- \frac{b^2}{z'^2}}\right\}}\frac{1}{\sqrt{1- \frac{b^2}{z'^2}}} \, ,
\end{eqnarray}
and hence
\begin{eqnarray}\label{eq:fdomain3}
\fl
f(z,w)  = \eta \rme^{-\eta z w}\int_{z\sqrt{1-w^2}}^{+\infty} \rmd z' \rho(z') \exp{\left\{-\eta \sqrt{z'^2-z^2(1-w^2)}\right\}}\frac{z'}{\sqrt{z'^2-z^2(1-w^2)}} \nonumber \\ \fl \phantom{f(z,w)  = } + \eta \rme^{-\eta z w} \int_{z\sqrt{1-w^2}}^z \rmd z' \rho(z') \exp{\left\{\eta \sqrt{z'^2-z^2(1-w^2)}\right\}}\frac{z'}{\sqrt{z'^2-z^2(1-w^2)}} \, .
\end{eqnarray}
\subsection{Domain 4: $w>0$ and $z\sqrt{1-w^2} = 1$}\label{sec:domain4}
The line defined by $z\sqrt{1-w^2} = 1$ with $w>0$ which is depicted in red in Fig.~\ref{fig:karact} plays a special role. Indeed, after each collision, the particle leaves the obstacle exactly along this line (except when it has flipped in the course of skidding) and it further keeps traveling along it until its self-propulsion tumbles. This strongly suggests that the bulk distribution function $f(z,w)$ displays a delta peak contribution along this characteristic line. To check this, we write
\begin{eqnarray}
f(z,w) = f_0(z,w) + \phi(z) \delta(z\sqrt{1-w^2}-1)\Theta(w) \, ,
\end{eqnarray}
with $f_0(z,w)$ a piece-wise continuous function (as $f_0$ might not be continuous when crossing the characteristic line $z\sqrt{1-w^2} = 1$ at $w>0$). By inserting the above expression in Eq.~\eqref{eq:FPbulk}, we get for $z > 1$
\begin{eqnarray}\fl
\left[\sqrt{1-\frac{1}{z^2}} \phi'(z) + \eta \phi(z) \right]\delta(z\sqrt{1-w^2}-1) + w \partial_z f_0 + \frac{1-w^2}{z}\partial_w f_0 + \eta (f_0 - \rho) = 0 \, ,
\end{eqnarray}
which in particular yields
\begin{eqnarray}
\phi(z) = \phi(1) \rme^{-\eta  \sqrt{z^2-1}} \, .
\end{eqnarray}
This shows that the existence of a singularity is equivalent to a non-vanishing $\phi(1)$. Remarkably, the value of $\phi(1)$ can be obtained from the flux-balance equation at the surface of the obstacle, Eq.~\eqref{eq:FPsurf}. By integrating it between $w = 0^{-}$ and $w = 0^{+}$ and using $\Gamma(w>0)=0$, we indeed obtain
\begin{eqnarray}
\Gamma(0^-) = \lim_{\epsilon \to 0} \int_{-\epsilon}^{+\epsilon} \rmd w \,\, w f(1,w) \, .
\end{eqnarray}
Hence 
\begin{eqnarray}
f(1,w) = \frac{\Gamma(0^-)}{w}\delta(w) + ... = \Gamma(0^-)\delta(\sqrt{1-w^2}-1) + ... \, , \end{eqnarray}
where the $...$ stands for something more regular than $\delta(w)/w$. We therefore get $\phi(1) = \Gamma(0^-)$ and thus
\begin{eqnarray}\label{eq:fdomain4}
\phi(z) = \Gamma(0^-)\rme^{-\eta  \sqrt{z^2-1}} \, .
\end{eqnarray}
We note that $\Gamma(0^-)$ is proportional to the rate of particles entering the red characteristics of Fig.~\ref{fig:karact} from the obstacle and its presence in Eq.~\eqref{eq:fdomain4} is therefore expected. The existence of a Dirac delta singularity in the distribution function, not only at the surface of the obstacle but also in the bulk of the $\left(w,z\right)$ plane, is a remarkable feature of this problem that, as far as we are aware, has not been pointed out in the literature. We believe this accumulation effect, that a soft-potential would smoothen out, is a generic feature of run-and-tumble particles around convex rigid obstacles. We furthermore expect that a continuous diffusive motion of the self-propulsion force (instead of a Poisson process as considered here), as is the case in other models of self-propelled particles like active Brownian particles \cite{solon2015active} and active Ornstein-Ulhenbeck ones \cite{martin2021statistical}, leads to a broadening of this singularity. 

\subsection{The self-consistency condition}
The bulk density $\rho(z)$ satisfies an integral equation that we obtain from Eq.~\eqref{eq:def_rho} by integrating Eqs.~\eqref{eq:fdomain1}-\eqref{eq:fdomain2}-\eqref{eq:fdomain3}-\eqref{eq:fdomain4} over the angular variable $w$. It takes the form of a linear integral equation involving a nonlocal kernel $\mathcal{L}$
\begin{eqnarray}
 \label{eq:self_consistent_density}
\rho(z)  = \rho_0(z) + \eta \mathcal{L}[\rho](z) \, ,
\end{eqnarray}
where
\begin{eqnarray} \label{eq:rhonot} \fl
    \rho_0(z) =   \, \eta \hat{\Gamma} \int_\frac{\sqrt{z^2-1}}{z}^1\frac{\rmd w}{2 W_{d-2}}(1-w^2)^{\frac{d-3}{2}} \frac{\exp{\left\{-\eta\left(zw- \sqrt{1-z^2(1-w^2)}\right)\right\}}}{\sqrt{1-z^2(1-w^2)}} \nonumber \\ \fl \phantom{ \rho_0(z) = } + \frac{\Gamma(0^-)}{2 W_{d-2}} \frac{\rme^{- \eta \sqrt{z^2-1}}}{\sqrt{z^2-1}} z^{2-d}  \, ,
\end{eqnarray}
and where the linear operator $\mathcal{L}$ is split into three nonlocal kernels
\begin{eqnarray} \label{eq:defL}
    \mathcal{L}[\rho](z) = \mathcal{L}_1[\rho](z) + \mathcal{L}_2[\rho](z) + \mathcal{L}_3[\rho](z) \, ,
\end{eqnarray}
with
\begin{eqnarray} \label{eq:defL1}\fl
    \mathcal{L}_1[\rho](z) = \int_1^{z} \rmd z' z' \rho(z')\int_\frac{\sqrt{z^2-1}}{z}^1\frac{\rmd w}{2 W_{d-2}}(1-w^2)^{\frac{d-3}{2}}\frac{\rme^{-\eta z w}  \exp{\left(\eta \sqrt{z'^2-z^2(1-w^2)}\right)} }{\sqrt{z'^2-z^2(1-w^2)}} \, , 
\end{eqnarray}
and
\begin{eqnarray} \label{eq:defL2}\fl
   \mathcal{L}_2[\rho](z) = 2 \int_{1}^z \rmd z'   z' \rho\left( z'\right) \, \int_{\frac{\sqrt{z^2-z'^2}}{z}}^{\frac{\sqrt{z^2-1}}{z}}\frac{\rmd w}{2 W_{d-2}} \left(1-w^2\right)^\frac{d-3}{2}\!\! \frac{\rme^{-\eta z w}\cosh{\left(\eta\sqrt{z'^2 - z^2(1-w^2)}\right)} }{\sqrt{z'^2-z^2(1-w^2)}} \, , \nonumber \\ 
\end{eqnarray}
and finally
\begin{eqnarray}\label{eq:defL3}\fl
   \mathcal{L}_3[\rho](z) = \int_z^{+\infty} \rmd z' z' \rho\left(z'\right) \, \int_{-1}^\frac{\sqrt{z^2-1}}{z} \frac{\rmd w}{2 W_{d-2}}\left(1-w^2\right)^{\frac{d-3}{2}}  \frac{\rme^{-\eta z w}  \exp{\left(-\eta \sqrt{z'^2-z^2(1-w^2)}\right)}}{\sqrt{z'^2 - z^2(1-w^2)}} \, . \nonumber \\ 
\end{eqnarray}
To the best of our knowledge, Eq.~\eqref{eq:self_consistent_density} can in general only be solved formally,
\begin{eqnarray}\label{eq:formal}
    \rho(z) = \sum_{n = 0}^{+\infty}\eta^n \mathcal{L}^{n}[\rho_0](z) \, .
\end{eqnarray}
Progress can nevertheless be made in some limiting cases. The first one that we study is the near obstacle regime $z - 1 \ll 1$ (equivalently $r-\sigma \ll \sigma$) at fixed $\eta = \sigma/\ell_p$. We then determine $\rho(z)$ in the highly ballistic limit $\eta\ll 1$ where the size of the obstacle is much smaller than the persistence length of the self-propelled particle and in a regime such that $\eta z \ll 1$ (equivalently $r \ll \ell_p$). Another case of interest are the corrections to the $\eta \to \infty$ equilibrium limit in which the persistence length is much smaller than the size of the obstacle. The first $1/\eta$ correction is equivalent to the case of one particle against a hard wall which was previously studied (in a two-dimensional geometry) in \cite{ezhilan2015distribution}. There the authors notably proved the existence of a delta-peak accumulation at the surface of the wall whose amplitude ($\hat{\Gamma}$ in our language) is proportional to $\eta^{-1}$. The situation is however very different from the generic $\eta$ result since in this limit particles have to tumble to leave the surface of the obstacle. In our language, this corresponds to $\Gamma(0^{-})=0$ (an equality that breaks down at finite values of $\eta$, see the numerical results in Fig.~\ref{fig:divergence_bulk}) and thus to the absence of a singularity in the bulk of the distribution function. In the ballistic regime, we prove in Sec.~\ref{sec:ballistic} that both $\Gamma(0^{-})$ and $\hat{\Gamma}$ have a finite non-zero limit that depends only on the dimension $d$.

\section{Behavior close to the obstacle: activity-induced attraction}\label{sec:vicinity}

\begin{figure}
  \centering
  \def\Y{2.0}
  \def\X{2.75}
  \begin{tikzpicture}
    \path (0,0)  node {\includegraphics[width=.45\columnwidth]{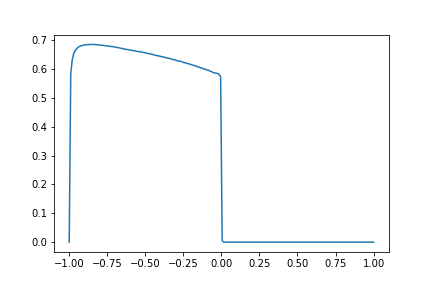}}
     (0.1,-2.4) node {$ w$}
     (-3.5,0) node[rotate=90] {$\Gamma(w)$};
    \path (7,0)  node {\includegraphics[width=.45\columnwidth]{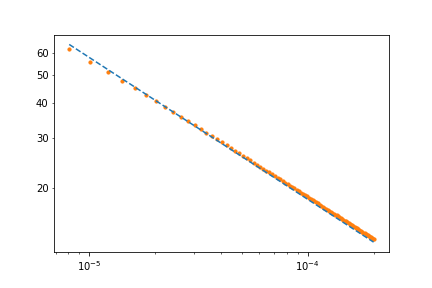}}
    (7,-2.3) node {$z^2-1$}
    (3.6,0) node[rotate=90] {$\rho(z)$};;
  \end{tikzpicture}
  \caption{Steady-state distribution of Eq.~\eqref{eq:EOM} for the hard sphere case at $\eta = 1$ in dimension $d = 2$.
  {\bf (Left)} Numerically measured surface distribution $\Gamma(w)$. It is non-vanishing for $w < 0$ only and yields $\Gamma(0^{-}) \simeq 0.57$. {\bf (Right)} Log-log plot of the bulk space density $\rho(z)$ as a function of $z^2-1$ in the vicinity of the obstacle. The dashed blue line is the theoretical prediction of Eq.~\eqref{eq:div_rhopasnot} with $2 W_0 = \pi$ and using the measured value of $\Gamma(0^{-})$ from the left panel. The orange dots correspond to numerical simulations. Parameters: $v = 1$, $\tau = 1$, $\sigma = 1$, $L = 100$, $T = 10^{10}$ with $T$ the total physical time.}
  \label{fig:divergence_bulk}
  \vspace{-0.10cm}
\end{figure}

In this section, we show that the bulk density $\rho(z)$ diverges close to the obstacle and relate exactly its diverging part to properties of the surface distribution function $\Gamma(w)$. In the following, we denote $h = z - 1$ and assume that $h \ll 1$. We recall the integral equation in Eq.~\eqref{eq:self_consistent_density} satisfied by the bulk density,
\begin{eqnarray}\label{eq:self_cons_recall}
\rho(z)  = \rho_0(z) + \eta \mathcal{L}[\rho](z) \, ,
\end{eqnarray}
with the linear operator $\mathcal{L}$ given in Eqs.~\eqref{eq:defL}-\eqref{eq:defL3}. As a first step, we investigate the small distance properties of the function $\rho_0(z)$. We find,
\begin{eqnarray}\label{eq:div_rhonot}
    \rho_0(1+h) = \frac{\Gamma(0^-)}{2 W_{d-2}}\frac{1}{\sqrt{2h}} - \frac{\eta \hat{\Gamma}}{4 W_{d-2}}\ln{h} + O(1) \, .
\end{eqnarray}
Indeed, the integral term of Eq.~\eqref{eq:rhonot} reads in the small $h$ limit,
\begin{eqnarray}
    \eta \hat{\Gamma} \int_\frac{\sqrt{z^2-1}}{z}^1\frac{\rmd w}{2 W_{d-2}}(1-w^2)^{\frac{d-3}{2}} \frac{\exp{\left\{-\eta\left(zw- \sqrt{1-z^2(1-w^2)}\right)\right\}}}{\sqrt{1-z^2(1-w^2)}} \nonumber \\ =  \, \frac{\eta \hat{\Gamma}}{4 W_{d-2} z^{d-2}} \int_0^{1} \rmd s\left(1-s\right)^{\frac{d-3}{2}}\frac{\exp{\left(\eta \left(\sqrt{s} - \sqrt{z^2-1+s}\right)\right)}}{\sqrt{s}\sqrt{z^2-1+s}} \, , \nonumber \\ =  \, - \frac{\eta \hat{\Gamma}}{4 W_{d-2}}\ln h + O(1) \, .
\end{eqnarray}
We find that Eq.~\eqref{eq:div_rhonot} gives the leading order behavior of the actual density field, \textit{i.e.}
\begin{eqnarray}\label{eq:div_rhopasnot}
    \rho(1+h) = \frac{\Gamma(0^-)}{2 W_{d-2}}\frac{1}{\sqrt{2h}} - \frac{\eta \hat{\Gamma}}{4 W_{d-2}}\ln{h} + O(1) \, .
\end{eqnarray}
Remarkably, the structure of the near-field density identified in Eq.~\eqref{eq:div_rhopasnot} is independent of the dimension $d$ for $d\geq 2$. It connects (without any free parameters) the divergences of the bulk distribution function on the obstacle to properties of the surface distribution function. We have simulated the dynamics Eq.~\eqref{eq:EOM} in dimension $d = 2$ and numerically measured $\Gamma(w)$ and $\rho(z)$. The simulation is performed in a spherical box of radius $L = 100 v_0 \tau$. The boundary condition is such that when the particle hits the outer boundary it is reflected with a random (inward) orientation. In Fig.~\ref{fig:divergence_bulk}, we plot the function $\Gamma(w)$ at $\eta = 1$ and infer the corresponding value of $\Gamma(0^{-})$ from it. We then show that the measured $\rho(z)$ indeed exhibits the square root divergence predicted by Eq.~\eqref{eq:div_rhopasnot}.

To show the validity of Eq.~\eqref{eq:div_rhopasnot}, our argument is the following. We first prove that $\mathcal{L}[\tilde{\rho}](1)$ is finite for any function $\tilde{\rho}(z)$ such that $\tilde{\rho}(1+h)\ln(h)$ is integrable at $h=0$. We then prove that $\rho(1+h)\ln(h)$ is integrable at $h=0$. This finally implies Eq.~\eqref{eq:div_rhopasnot} from Eq.~\eqref{eq:self_cons_recall}. We study the operators $\mathcal{L}_{1,2,3}$ separately. We start by noting that upon a change of variables we have
\begin{eqnarray}
\fl
    \mathcal{L}_1[\tilde{\rho}](1+h) = \frac{(1+h)h}{4W_{d-2}}\int_0^1 \rmd u \left(1+u h\right)\tilde{\rho}\left(1+u h\right) g_1\left(1+u h, 1 + h\right) \, ,
\end{eqnarray}
with 
\begin{eqnarray}\fl
    g_1(z',z) = \int_0^{1} \rmd s \frac{\left(1-s\right)^{\frac{d-3}{2}}}{\sqrt{z^2-1+s}}\frac{\exp{\left(-\eta\sqrt{z^2-1+s} + \eta \sqrt{z'^2-1+z^2 s}\right)}}{z^2\sqrt{z'^2-1+s}} \, .
\end{eqnarray}
For $h \ll 1$, we obtain
\begin{eqnarray}
    g_1\left(1+u h, 1 + h\right)  \simeq \int_0^1 \rmd s \frac{1}{\sqrt{s\left(s + 2h(1+u)\right)}} \, , \nonumber \\ \phantom{g_1\left(1+u h, 1 + h\right)} \simeq - \ln h \, .
\end{eqnarray}
Therefore,
\begin{eqnarray}\label{eq:vanishL1}
     \mathcal{L}_1[\tilde{\rho}](1+h) \simeq - \frac{1}{4 W_{d-2}} h \ln h \int_0^1 \rmd u \,  \tilde{\rho}\left(1+ u h\right) \underset{h \to 0}{\to} 0 \, ,
\end{eqnarray}
as the function $\tilde{\rho}\left(1+ h\right)$ is assumed to be integrable at $h=0$. We proceed accordingly for $\mathcal{L}_2$. We have,
\begin{eqnarray}
    \mathcal{L}_2[\tilde{\rho}](1+h) = 2  h \int_0^1 \rmd u \left(1 + u h\right) \tilde{\rho}\left(1+u h\right) g_2\left(1+u h, 1 + h\right) \, ,
\end{eqnarray}
with
\begin{eqnarray}\fl
g_2(z',z) = \int_{\frac{\sqrt{z^2-z'^2}}{z}}^{\frac{\sqrt{z^2-1}}{z}}\frac{\rmd w}{2 W_{d-2}} \left(1-w^2\right)^\frac{d-3}{2}\!\!\!\!\!\! \frac{\exp{\left(-\eta z w\right)}}{\sqrt{z'^2-z^2(1-w^2)}} \cosh{\left(\eta\sqrt{z'^2 - z^2(1-w^2)}\right)} \, .
\end{eqnarray}
For $h \ll 1$, we obtain
\begin{eqnarray}
    g_2\left(1+u h, 1 + h\right)  \simeq \int_{\sqrt{2h(1-u)}}^{\sqrt{2h}} \frac{\rmd w}{2W_{d-2}} \frac{1}{\sqrt{2h(u-1) + w^2}} \, , \\  \phantom{g_2\left(1+u h, 1 + h\right)} \simeq {\rm arctanh} \left(\sqrt{u}\right) \, .
\end{eqnarray}
Hence we get,
\begin{eqnarray}\label{eq:vanishL2}
    \mathcal{L}_2[\tilde{\rho}](1+h) \simeq 2 h \int_0^1 \rmd u \, \tilde{\rho}\left(1+u h\right){\rm arctanh}\left(\sqrt{u}\right) \underset{h \to 0}{\to} 0 \, ,
\end{eqnarray}
by integrability of $\tilde{\rho}$. Lastly, we show that if $\tilde{\rho}(1+h)\ln(h)$ is integrable at $h=0$ then $\mathcal{L}_3[\tilde{\rho}](1)$ is finite. From Eq.~\eqref{eq:defL3}, the latter yields a finite result if 
\begin{eqnarray}
    \tilde{\rho}\left(z'\right) \int_{-1}^0 \frac{\rmd w}{2 W_{d-2}}\left(1-w^2\right)^{\frac{d-3}{2}} \frac{\exp{\left(\eta w- \eta \sqrt{z'^2-1+w^2)}\right)}}{\sqrt{z'^2-1+w^2}} \, ,
\label{eq:integrability}
\end{eqnarray}
is integrable at $z' = 1$. Furthermore, for $h' = \sqrt{z'^2 - 1} \ll 1$, we have,
\begin{eqnarray}\fl\label{eq:L3lim}
      \int_{-1}^0 \frac{\rmd w}{2 W_{d-2}}\left(1-w^2\right)^{\frac{d-3}{2}} \frac{\exp{\left(\eta w- \eta \sqrt{z'^2-1+w^2}\right)}}{\sqrt{z'^2-1+w^2}} \simeq  \frac{-1}{4W_{d-2}}\ln h' \, .
\end{eqnarray}
Therefore, $\mathcal{L}_3[\tilde{\rho}](1)$ is indeed finite if $\tilde{\rho}\left(1+h\right) \ln(h)$ is integrable at $h = 0$. 

Then, if $\rho\left(1+h\right) \ln(h)$ is integrable, $\mathcal{L}[\rho](1)$ is finite and we recover Eq.~\eqref{eq:div_rhopasnot} from Eq.~\eqref{eq:self_cons_recall}. We now assume that $\rho\left(1+h\right) \ln(h)$ is not integrable at $h = 0$ only to prove that it is not possible at the end. If that is the case, we must have close to $h=0$
\begin{equation}\label{eq:equiv1}
\rho(1+h) \sim \mathcal{L}[\rho](1+h) \, ,
\end{equation}
since $\rho_0(1+h)\ln(h)$ is integrable at $h=0$. By virtue of Eqs.~\eqref{eq:vanishL1}-\eqref{eq:vanishL2} and Eqs.~\eqref{eq:integrability}-\eqref{eq:L3lim}, Eq.\eqref{eq:equiv1} becomes
\begin{equation}\label{eq:equiv2}
\rho(1+h) \sim \mathcal{L}_3[\rho](1+h) \sim \int_h \rmd h' \rho(1+h')\ln(h') \,.
\end{equation}
However, as we show in \ref{app:bound}, since $\rho(1+h)$ is integrable at $h=0$, the integral term in the right-hand side of Eq.\eqref{eq:equiv2} grows at most as $\ln h$ close to $h=0$, thus contradicting the initial assumption that $\rho(1+h)\ln(h)$ is not integrable at $h=0$. Therefore, $\rho(1+h)\ln(h)$ is integrable at $h=0$ which in turns implies Eq.~\eqref{eq:div_rhopasnot}. 

\section{Going highly ballistic}\label{sec:ballistic}
\subsection{Steady-state distribution at $\eta = 0$}\label{sec:limit}
\begin{figure}
  \centering
  \begin{tikzpicture}
    \path (0,0)  node {\includegraphics[width=.45\columnwidth]{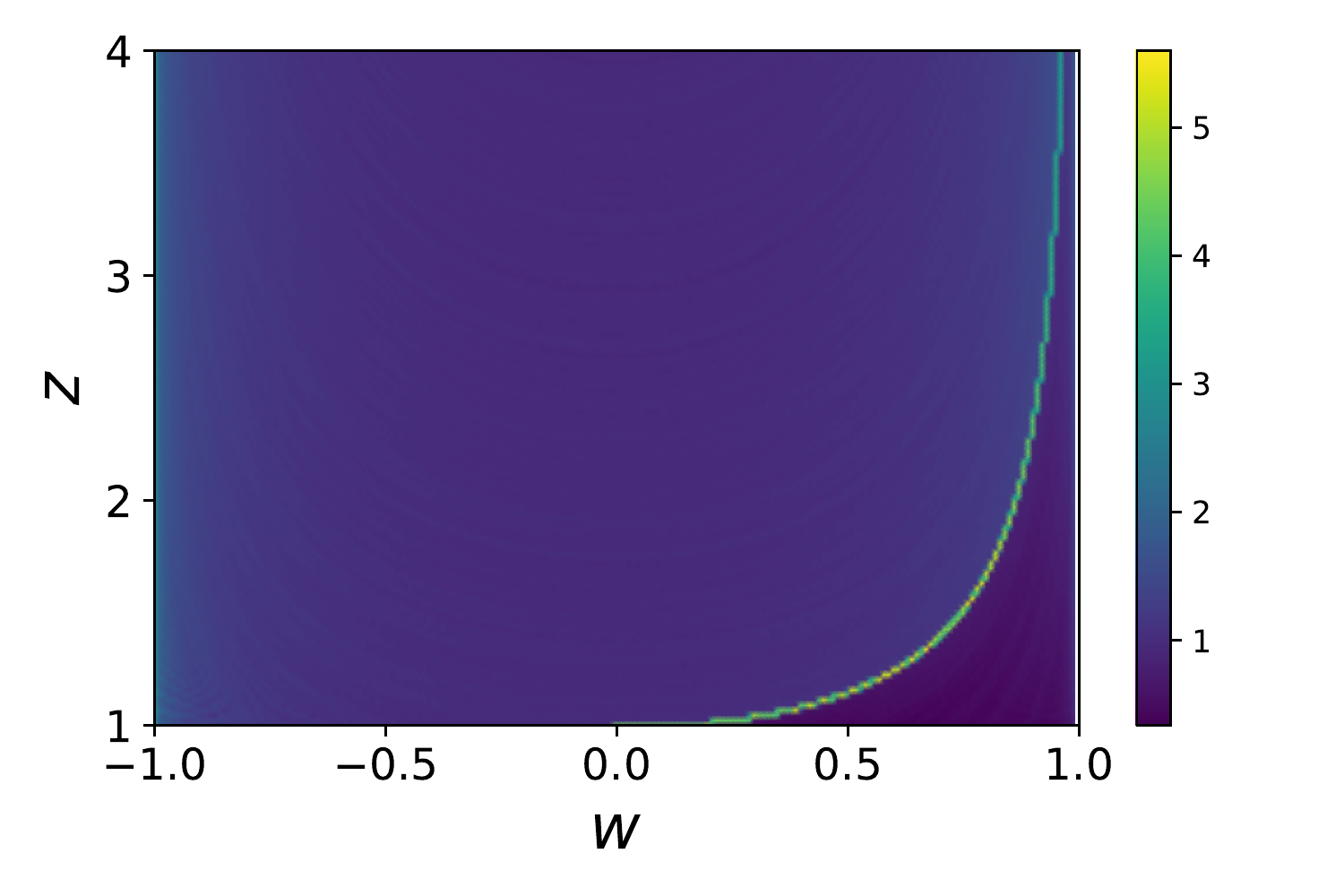}};
  \end{tikzpicture}
  \caption{Heatmap of the bulk distribution function $f(z,w)$. It shows two regions of nearly homogeneous densities, one at higher density for $w<0$ and $w>0$ with $z\sqrt{1-w^2} \geq 1$ and one at lower density for $w>0$ with $z\sqrt{1-w^2}<1$. The two domains are separated by a line of high density. Parameters: $v = 1$, $\tau = 1$, $L = 10$, $\eta = 0.1$ and $T = 10^{7}$ with $T$ the total physical time.}
  \label{fig:heatmap}
  \vspace{-0.10cm}
\end{figure}

We now study the solution of Eq.~\eqref{eq:self_consistent_density} in the highly ballistic limit $\eta = \sigma/\ell_p \to 0$ and we begin by stating our results. When $\eta = 0$, the surface distribution function is uniform over $w\in[-1,0]$,
\begin{eqnarray}\label{eq:resultgamma}
    \Gamma(w) = \frac{1}{d-1} \Theta(-w) \, ,
\end{eqnarray}
and the amplitude of the delta peak accumulation therefore reads 
\begin{eqnarray}\label{eq:resultgammahat}
    \hat{\Gamma} = \frac{1}{2(d-1)} \, .
\end{eqnarray}
Furthermore, for any $z$ fixed (hence for $r \ll \ell_p$), the particle density is given by
\begin{eqnarray}\label{eq:resultrho}
\fl    \rho(z) = \Theta(z-1)\left[\frac{1}{2} + \frac{\sqrt{z^2-1}}{2 W_{d-2} z}{}_2 F_1\left(\frac{1}{2},\frac{3-d}{2},\frac{3}{2};1-\frac{1}{z^2}\right) + \frac{1}{2 W_{d-2} (d-1)}\frac{z^{2-d}}{\sqrt{z^2-1}} \right] \nonumber \\ + \frac{1}{2(d-1)} \delta(z-1)  \, ,
\end{eqnarray}
where we recall that the boundary conditions are such that $\rho(\infty)=1$. We introduce  the density offset created by the obstacle defined by $K(z)=\rho(z)-1$. Away from the obstacle, we find the power-law decay 
\begin{eqnarray} \label{eq:equivz}
   K(z) \sim \frac{1}{2 W_{d-2} (d^2-1)}\frac{1}{z^{d+1}}\,,
\end{eqnarray}
which holds as long as $\eta z\ll 1$ (equivalently $r \ll \ell_p$)  after which the density modulations are exponentially suppressed. Importantly, the above expression is integrable at $z\to\infty$ for any $d\geq 2$. From there we obtain the stationary distribution in the $(z,w)$ plane. In domain 1 defined by $w>0$ and $z\sqrt{1-w^2} < 1$, Eq.~\eqref{eq:fdomain1} reduces to 
\begin{eqnarray}
    f(z,w) = 0 \, .
\end{eqnarray}
as $\eta \to 0$. In domain 2 defined by $w < 0$, Eq.~\eqref{eq:fdomain2} is given in this limit by
\begin{eqnarray}\label{eq:solwneg}
\fl
f(z,w)  = 1 + \eta \rme^{-\eta z w} \int_z^{+\infty} \rmd z' K(z') \exp{\left\{-\eta \sqrt{z'^2- z^2(1-w^2)}\right\}}\frac{z'}{\sqrt{z'^2- z^2(1-w^2)}} \, , \\ \fl  \phantom{f(z,w)} = 1 \, .
\end{eqnarray}
Finally, in domain 3 defined by $w>0$ and $z\sqrt{1-w^2} > 1$ we obtain accordingly from Eq.~\eqref{eq:fdomain3},
\begin{eqnarray}\label{eq:fdomain3sol}
f(z,w)  = 1  \, .
\end{eqnarray}
Therefore, the full bulk distribution reads
\begin{eqnarray}\label{eq:result_final_f}
 \fl   f(z,w) = \Theta(-w) + \Theta(w)\left[\Theta(z\sqrt{1-w^2}-1) + \frac{1}{d-1} \delta(z\sqrt{1-w^2}-1)\right] \, .
\end{eqnarray}
In Fig.~\ref{fig:heatmap}, we show a heatmap of the bulk distribution function $f(z,w)$ at $\eta = 0.1$. It exhibits two regions of nearly homogeneous densities, one at higher density for $w<0$ and $w>0$ with $z\sqrt{1-w^2} \geq 1$ and one at lower density for $w>0$ with $z\sqrt{1-w^2}<1$. The two domains are separated by a line of high density located at the locus of the delta peak singularity of Eq.~\eqref{eq:result_final_f}.

We now prove Eqs.~\eqref{eq:resultgamma}-\eqref{eq:resultrho} and then we give the expressions of the next-to-leading order $O(\eta)$ corrections to the amplitude of the delta peak showing that the latter is a decreasing function of $\eta$. Tumbling thus reduces the fraction of time spent in contact with the obstacle. It follows from Eq.~\eqref{eq:self_consistent_density} that, for $z>1$, $K(z)$ is a solution of the integral equation
\begin{eqnarray}\label{eq:self_cons_K}
    K(z) = K_0(z) + \eta \mathcal{L}[K](z) \, ,
\end{eqnarray}
with $\mathcal{L}$ defined in Eqs.~\eqref{eq:defL}-\eqref{eq:defL3} and where
\begin{eqnarray}
\fl \label{eq:expK0}
K_0(z) = \rho_0(z) - \int_{\frac{\sqrt{z^2-1}}{z}}^{1}\frac{\rmd w}{2 W_{d-2}} \left(1-w^2\right)^{\frac{d-3}{2}}\exp{\left(-\eta z w + \eta \sqrt{1 - z^2(1-w^2)}\right)} \, .
\end{eqnarray}
So far, $\rho_0(z)$, which is defined in Eq.~\eqref{eq:rhonot}, explicitly depends on $\Gamma(0^-)$ and $\hat{\Gamma}$. In order to obtain Eq.~\eqref{eq:resultgamma} and a closed equation for $K(z)$ we start by defining $K_{\eta = 0}(z) = \lim_{\eta\to 0} K(z)$ and, consistently with Eq.~\eqref{eq:equivz}, we assume that $K_{\eta = 0}(z)$ is integrable as $z\to\infty$. Under this assumption, in the $\eta \to 0$ limit, Eq.~\eqref{eq:fdomain2} can thus be rewritten as
\begin{eqnarray}
\fl f(1,w<0)  = 1 + \eta \rme^{-\eta w} \int_1^{+\infty} \rmd z' \exp{\left\{-\eta \sqrt{z'^2-(1-w^2)}\right\}}\frac{ K(z')  z'}{\sqrt{z'^2-(1-w^2)}} \, , \nonumber \\ \fl \phantom{f(1,w<0)}  = 1 + \eta \int_1^{+\infty} \rmd z' \frac{ K_{\eta = 0}(z')  z'}{\sqrt{z'^2-(1-w^2)}} + o(\eta) \,.
\label{eq:fdomain2bis}
\end{eqnarray}
Together with Eq.~\eqref{eq:FPsurf}, the above equation therefore yields Eq.~\eqref{eq:resultgamma}, 
\begin{eqnarray}
\Gamma(w) = \frac{1}{d-1}\Theta(-w) + \eta \Gamma_1(w) + o(\eta) \,,
\label{eq:proofGamma}
\end{eqnarray}
where the correcting term $\Gamma_1(w)$ is, for $w<0$, a solution of
\begin{eqnarray}\label{eq:first_order_Gamma}
\fl
\Gamma_1'(w) - \frac{w}{1-w^2}(d-1) \Gamma_1(w) = - \frac{w}{1-w^2}\int_1^{+\infty} \rmd z' \frac{ K_{\eta = 0}(z')  z'}{\sqrt{z'^2-(1-w^2)}} - \frac{1}{2(d-1)(1-w^2)} \,. \nonumber \\ 
\end{eqnarray}
We defer to Sec.~\ref{sec:corr} the study of the small $\eta$ corrections but we note that Eq.~\eqref{eq:resultgamma} follows from Eq.~\eqref{eq:proofGamma} obtained under the sole assumption that $K_{\eta = 0}(z)$ is integrable as $z\to\infty$. To leading order in $\eta$, we thus obtain from Eq.~\eqref{eq:expK0},
\begin{eqnarray} \label{eq:soleta0}
\fl \lim_{\eta \to 0} K_{0}(z) = \left[-\frac{1}{2} + \frac{\sqrt{z^2-1}}{2 W_{d-2} r}{}_2 F_1\left(\frac{1}{2},\frac{3-d}{2},\frac{3}{2};1-\frac{1}{z^2}\right) + \frac{1}{2 W_{d-2} (d-1)}\frac{z^{2-d}}{\sqrt{z^2-1}} \right] \, ,
\end{eqnarray}
which is integrable for $z\to\infty$. The rest of the proof follows from the fact that if $K_{\eta = 0}(z)$ is integrable at $z\to\infty$ then $\lim_{\eta\to0} \mathcal{L}[K](z)$ exists. We first notice that
\begin{eqnarray}
    \underset{\eta \to 0}{\lim} \mathcal{L}_1[K](z) = \int_1^z \rmd z' \, z' \, K_{\eta = 0}\left(z'\right) \int_{\frac{\sqrt{z^2-1}}{z}}^1 \frac{\rmd w}{2 W_{d-2}}  \frac{\left(1-w^2\right)^{\frac{d-3}{2}}}{\sqrt{z'^2 - z^2\left(1-w^2\right)}} \, .
\end{eqnarray}
The above integral is indeed well-defined as 
\begin{eqnarray}
    \int_{\frac{z^2-1}{z}}^1 \frac{\rmd w}{2 W_{d-2}}  \frac{\left(1-w^2\right)^{\frac{d-3}{2}}}{\sqrt{z'^2 - z^2\left(1-w^2\right)}} \, ,
\end{eqnarray}
remains finite at $z' = 1$ and $K_{\eta = 0}\left(z'\right)$ is integrable at $z' = 1$. Accordingly, the same holds for $\mathcal{L}_2$ since
\begin{eqnarray}\label{eq:L2neglect}
   \fl \underset{\eta \to 0}{\lim} \mathcal{L}_2[K](z) =  2 \int_{1}^z \rmd z'   z' K_{\eta = 0}\left( z'\right) \, \int_{\frac{\sqrt{z^2-z'^2}}{z}}^{\frac{\sqrt{z^2-1}}{z}}\frac{\rmd w}{2 W_{d-2}}  \frac{\left(1-w^2\right)^\frac{d-3}{2}}{\sqrt{z'^2-z^2(1-w^2)}} \, .
\end{eqnarray}
We now turn to the study of $\mathcal{L}_3$. We recall Eq.~\eqref{eq:defL3},
\begin{eqnarray}
 \fl  \mathcal{L}_3[K](z) = \int_z^{+\infty} \rmd z' z' K\left(z'\right) \, \int_{-1}^\frac{\sqrt{z^2-1}}{z}  \frac{\rmd w}{2 W_{d-2}} \left(1-w^2\right)^{\frac{d-3}{2}}  \times \\  \!\!\!\!\dots \times \frac{\exp{\left(-\eta z w- \eta \sqrt{z'^2-z^2(1-w^2)}\right)}}{\sqrt{z'^2 - z^2(1-w^2)}} \, .
\end{eqnarray}
from which it appears that if $ K_{\eta=0}\left(z'\right)$ is integrable at $+\infty$, then
\begin{eqnarray}\label{eq:equivL3}
    \underset{\eta \to 0}{\lim} \,  \mathcal{L}_3[K](z) \,\,\, \rm{exists.}
\end{eqnarray}
The integrability of $\lim_{\eta \to 0} K_{0}(z)$ for $z\to\infty$ therefore guarantees that for $z>1$
\begin{eqnarray}\label{eq:proff}
    K(z) = K_0(z) + \eta \lim_{\eta \to 0} \mathcal{L}[K_{0}](z) + o(\eta) \,,
\end{eqnarray}
which self-consistently proves the integrability of $K_{\eta = 0}(z)$ which was our starting assumption. Equation \eqref{eq:resultrho} then follows immediately. The result in Eq.~\eqref{eq:resultrho} is independent of the persistence length $\ell_p$ of the RTP. It only depends on $\sigma$ through the rescaling $r = \sigma z$.

Two interesting results follow. First, the main results of this section, notably Eqs. ~\eqref{eq:resultgamma}-\eqref{eq:resultrho}-\eqref{eq:result_final_f}, also hold for different classes of self-propelled particles \cite{solon2015active,martin2021statistical}. For active Brownian particles whose self-propulsion has norm $v_0$ and correlation time $\tau$, the steady state distribution function also depends only on $z$ and $w$ and obeys
\begin{eqnarray} \fl
   0 =  - \left(w \partial_r P(z,w) + \frac{1-w^2}{z}\partial_w P(z,w)\right) + \frac{1}{z^{d-1}}\partial_z\left(z^{d-1} P(z,w) \partial_z V(z) \right) \nonumber \\  + \eta \mathcal{R}P(z,w)\, .
\label{eq:FP_gen}
\end{eqnarray} 
with $\mathcal{R}$ an operator independent of $\eta$ accounting for the reorientation of the active self-propulsion, see \textit{e.g.} \cite{solon2015active} for a comparative review on run-and-tumble and active Brownian particles. In the limit $\eta\to 0$, the subtleties associated to the details of the operator $ \mathcal{R}$ are therefore suppressed and our results derived for run-and-tumble particles are expected to hold. We also conjecture that our results also hold for active Ornstein-Ulhenbeck particles \cite{martin2021statistical}, which may come more as a surprise since the norm of the active self-propulsion fluctuates in this model. There, the steady-state distribution function indeed depends on $(z,w)$ and on $v$ the self-propulsion velocity. Here we denote $v_0 = \left\langle v \right\rangle$ its average and introduce as before $\eta = \sigma / (v_0\tau)$. The steady-state distribution obeys 
\begin{eqnarray} \fl
   0 =  - \left(w \partial_r P(z,w,v) + \frac{1-w^2}{z}\partial_w P(z,w,v)\right) + \frac{1}{z^{d-1}}\partial_z\left(z^{d-1} P(z,w,v) \frac{v_0}{v} \partial_z V(z) \right) \nonumber \\  + \frac{v_0}{v} \eta \mathcal{R}P(z,w,v)\, .
\label{eq:FP_AOUP}
\end{eqnarray} 
with $\mathcal{R}$ another operator independent of $\eta$ accounting for the reorientation of the active self-propulsion. In the $\eta\to 0$ limit, and in the hard-sphere limit where the prefactor $\frac{v_0}{v}$ in front of the gradient of the potential becomes irrelevant, it follows that $v$ and $(z,w)$ decouple in the steady-state and that the results obtained in this section for run-and-tumble particles provide the correct marginal in the $(z,w)$ space of the steady-state distribution of a highly-ballistic active Ornstein-Ulhenbeck particle interacting with a hard-sphere obstacle.

Additionally, Eq.~\eqref{eq:resultrho} also corresponds to the steady-state distribution function of two highly-ballistic RTPs interacting via hard-core repulsion -- or equivalently the low-density pair-distribution function of a fluid of highly ballistic hard-core RTPs -- with $\sigma$ representing the diameter of a single particle. Indeed, the relative separation $\bi{r} = \bi{r}_1 - \bi{r}_2$ of the two run-and-tumble particles, with self-propulsions along $\bi{u}_1$ and $\bi{u}_2$, evolves according to
\begin{equation}
\frac{\rmd \bi{r}}{\rmd t} = v_0 ||\bi{u}_1 - \bi{u}_2|| \bi{n} - 2 \bi{\nabla} V(\bi{r}) \,.
\end{equation} 
with the relative self-propulsion direction
\begin{equation}
\bi{n} = \frac{\bi{u}_1 - \bi{u}_2}{||\bi{u}_1 - \bi{u}_2||} \,.
\end{equation}
At large distances, $\bi{n}$ is isotropically distributed on the unit sphere. Therefore, in the highly ballistic limit, where the self-propulsion $\bi{n}$ is only set by boundary conditions, the problem maps to that of a single run-and-tumble particle studied in this work. The highly-ballistic steady-state distribution function being independent of the persistence length $\ell_p$, Eq.~\eqref{eq:resultrho} also holds for a pair of highly-ballistic particles.

\subsection{Corrections to the ballistic limit}\label{sec:corr}

\begin{figure}
  \centering
  \def\Y{2.0}
  \def\X{2.75}
  \begin{tikzpicture}
    \path (0,0)  node {\includegraphics[width=.45\columnwidth]{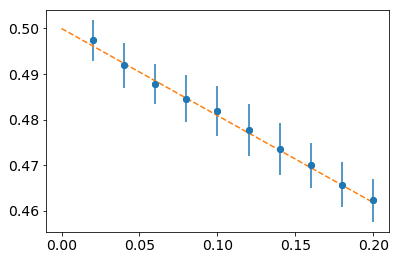}}
     (0.3,-2.3) node {$\eta$}
     (-3.8,0.3) node[rotate=90] {$\hat{\Gamma}$};
  \end{tikzpicture}
  \caption{Amplitude of the delta peak accumulation at the surface of the obstacle $\hat{\Gamma}$ in dimension $d=2$ as a function of the ratio $\eta$ of the obstacle size over the persistence length of the RTP. The dashed orange line is the theoretical prediction of Eq.~\eqref{eq:decayd2}. The blue dots correspond to numerical simulations at fixed persistence length with varying obstacle sizes. Parameters: $v = 1$, $\tau = 1$, $L = 20$, $T = 10^{10}$ with $T$ the total physical time.}
  \label{fig:decayd2}
  \vspace{-0.10cm}
\end{figure}

Equation~\eqref{eq:first_order_Gamma} gives the first order corrections to the ballistic limit of the surface term of the steady-state distribution function. Imposing integrability at $w=-1$, we obtain in dimension $d = 2$,
\begin{eqnarray}
\fl
\Gamma_1(w) = \frac{-(2 + \pi w) \sqrt{1 - w^2} + \pi (-\pi + {\rm arccos}(w)) - 
 2 w^2 {\rm arctanh}(\sqrt{1 - w^2})}{4 \pi \sqrt{1 - w^2}} \,,
\end{eqnarray}
thus yielding
\begin{eqnarray} \label{eq:decayd2}
\hat{\Gamma} = \frac{1}{2} - \eta \frac{4 C - 2 + \pi}{8 \pi } + o(\eta) \,,
\end{eqnarray}
with $C \simeq 0.92$ Catalan's constant. We numerically confirmed the decay of the amplitude of the surface term predicted by Eq.~\eqref{eq:decayd2} as shown in Fig.~\ref{fig:decayd2}.
In dimension $d = 3$, Eq.~\eqref{eq:first_order_Gamma} allows to express $\Gamma_1(w)$ in terms of $K$ and $E$, which are the complete elliptic integral of the first and second kind, respectively,
\begin{eqnarray}
\fl
\Gamma_1(w) = \frac{6 w^2 K\left(1-w^2\right)+3 \pi  \left(w^2-1\right) w \, _2F_1\left(-\frac{1}{2},\frac{3}{2};2;1-\frac{1}{w^2}\right)-4 w^3-6 E\left(1-w^2\right)+6 w+2}{24 \left(w^2-1\right)}  \,. \nonumber \\ 
\end{eqnarray}
We therefore obtain
\begin{eqnarray}
\hat{\Gamma} = \frac{1}{4} - c \eta + o(\eta) \,,
\end{eqnarray}
with the constant $c$ given by
\begin{eqnarray}
c = - \int_{-1}^0 \frac{\rmd w}{2} \Gamma_1(w) \simeq 0.053\,.
\end{eqnarray} 

\section{The effect of hydrodynamic interactions}\label{sec:Hydro}
In the above, we studied the steady state distribution function of a run-and-tumble particle around a fixed spherical obstacle. In the highly ballistic limit, the latter matches the steady-state distribution of two run-and-tumble particles with hard-core repulsion. In the present section, we consider the motion of two run-and-tumble particles interacting via hydrodynamic interactions carried by a Stokesian fluid of viscosity $\mu$ and a repulsive hard-core potential in dimension $d\geq3$. We deal with colloids driven by an external force and endowed with a no-slip boundary condition at their surface, as can be implemented by trapping them in mobile optical traps \cite{berner2018oscillating}. In the overdamped limit, the equations of motion express a balance between the propulsion force, the potential interaction between the particles and the force exerted by the surrounding fluid on each of them. This results in a effective position dependent mobility for each particle given, to leading order in their relative separation, by the Oseen tensor. The Oseen tensor is the first term of an infinite series in powers of $\sigma/r$ which does not fully capture near-field phenomenology by overestimating the mobility in that regime \cite{guazzelli2011physical}. Retaining the Oseen tensor only, as we shall now implement, can be seen as a first step towards understanding the role of hydronamics. As in the dry case, we find that in this approximation the steady-state distribution displays a singular-at-contact delta-peak accumulation. Furthermore, we show that the main features of the highly-ballistic bulk density obtained in the previous section, \textit{i.e.} a square-root divergence close to contact and a power-law decay with exponent $-4$ in $d=3$ at large distances (but on scales smaller than the persistence length) survive. The latter feature, the $r^{-4}$ decay, is likely to be robust upon expanding beyond the Oseen approximation. Our study also predicts a reduced sticking coefficient compared to the dry case. Refined results on the effect of hydrodynamic interactions could be obtained by using the Rotne-Prager-Yamakawa \cite{rotne1969variational, yamakawa1970transport} approximation and subsequent expansions of the mobility tensor as derived in \cite{felderhof1977hydrodynamic}. For completeness we provide in \ref{app:Oseen} a derivation of the equations of motion used in this section. Upon explicitly reintroducing a mobility $\omega$ for the purpose of the discussion, these can be written as
\begin{equation}
\label{eq:EOM_hydro}
\fl
\cases{0 = f_0 \bi{u}_1  - \bnabla_\bi{r} V(\bi{r}) - \omega\left[\frac{\rmd \bi{r}_1}{\rmd t} - A(r) \frac{\rmd \bi{r}_2}{\rmd t} - (d-2) A(r) \frac{1}{r^2}\left(\bi{r} \cdot \frac{\rmd \bi{r}_2}{\rmd t}\right)\bi{r} \right]\,,\\
0 = f_0 \bi{u}_2 +  \bnabla_\bi{r} V(\bi{r}) - \omega\left[\frac{\rmd \bi{r}_2}{\rmd t} - A(r) \frac{\rmd \bi{r}_1}{\rmd t} - (d-2) A(r) \frac{1}{r^2}\left(\bi{r} \cdot \frac{\rmd \bi{r}_1}{\rmd t}\right)\bi{r} \right]\,,\\}
\end{equation}
with $\bi{r} = \bi{r}_1 - \bi{r}_2$ and
\begin{eqnarray}
        \omega = \mu \frac{ d(d-2)}{d-1} \frac{\Omega_d\sigma^{d-2}}{2^{d-2}} \nonumber \,, \\
        A(r) = \frac{d}{2(d-1)}\left(\frac{\sigma}{2r}\right)^{d-2} \,,
\end{eqnarray}
yielding
\begin{eqnarray}
    \left[1 + A(r) + (d-2) \frac{A(r)}{r^2}\left(\bi{r} \cdot \right)\bi{r}\right]\frac{\rmd \bi{r}}{\rmd t} = \frac{f_0}{\omega}\left(\bi{u}_1 - \bi{u}_2\right) - \frac{2}{\omega}\bnabla_\bi{r} V(\bi{r})\,.
\end{eqnarray}
To first order in $r\gg\sigma$, we can easily invert this equation and obtain
\begin{eqnarray}
    \frac{\rmd r^\alpha}{\rmd t} & = M^{\alpha\beta}\left(v_0\left(u_1^\beta-u_2^\beta\right) - 2 \partial_\beta V(r)\right) \,,
\end{eqnarray}
with $v_0 = f_0/\omega$ and where we have rescaled the potential $V(\bi{r}) \rightarrow V(\bi{r})/\omega$. The tensor $M^{\alpha\beta}$ is given by
\begin{eqnarray}
     M^{\alpha \beta} = \delta^{\alpha\beta}(1-A(r)) - (d-2) A(r) \frac{r^\alpha r^\beta}{r^2}\,,
\end{eqnarray}
which allows us to rewrite the equation of motion as
\begin{eqnarray}
\fl   \frac{\rmd \bi{r}}{\rmd t} = \left(1-A(r)\right)v_0\left(\bi{u}_1-\bi{u}_2\right) -v_0(d-2)A(r)\left[\left(\bi{u}_1-\bi{u}_2\right)\cdot\hat{\bi{r}}\right]\hat{\bi{r}} \nonumber \\ - 2 \left(1 - (d-1)A(r)\right)\bnabla_\bi{r} V(\bi{r})\,.
\end{eqnarray}
Several comments are in order. First the amplitude of the pairwise forces is reduced due to hydrodynamic interactions. Second, the effective propulsion velocity is also reduced and it becomes space dependent: the velocity is lower as the particles get closer. Third, there is also an extra force along $\hat{\bi{r}}$. Note that it is repulsive if $\left(\bi{u}_1-\bi{u}_2\right)\cdot\hat{\bi{r}} < 0$ and attractive if $\left(\bi{u}_1-\bi{u}_2\right)\cdot\hat{\bi{r}} > 0$. We will show that this results in a reduced delta-peak attraction with respect to the dry case. The stationary state distribution function $g_0(\bi{r};\bi{u}_1,\bi{u}_2)$ is the solution of
\begin{eqnarray}
    0 = - \partial_\alpha \left(g_0 M^{\alpha\beta}\left(v_0\left(\bi{u}_1^\beta - \bi{u}_2^\beta\right) - 2 \partial^\beta V(\bi{r})\right)\right) + \mathcal{R}g_0 \,,
\end{eqnarray}
where the operator $\mathcal{R}$, which is negligible in the highly ballistic limit, accounts for the reorientation of the active forces. In this limit, when the pair-potential is taken to be of the hard-sphere type, the bulk distribution function $f$ obeys
\begin{eqnarray}\label{eq:charact_hydro}
    w \left[1-(d-1)A(z)\right]\partial_z f + \left(1-A(z)\right)\frac{1-w^2}{z}\partial_w f = 0 \,.
\end{eqnarray}
with $w$ the cosine of the angle between $\bi{u}_1 - \bi{u}_2$ and $\hat{\bi{r}}$ and $z = r/\sigma$. Furthermore, the surface term satisfies 
\begin{eqnarray}\label{eq:surf8hydro}
  \fl  \Gamma'(w) - (d-1) \frac{w}{1-w^2} \Gamma(w) & =  - f(\sigma,w) \frac{w}{1-w^2} \left[1 - \frac{(d-2)A(\sigma)}{1-A(\sigma)}\right] \,, \\ & = - (d-1) f(1,w) \frac{w}{1-w^2} \left(\frac{1-d2^{1-d}}{d-1 - d 2^{1-d}}\right) \,.
\end{eqnarray}
for $w<0$ and $\Gamma(w>0)=0$. In the highly ballistic limit, the bulk distribution function $f$ is constant along the characteristics of Eq.~\eqref{eq:charact_hydro} which read
\begin{eqnarray}
    z \sqrt{1-w^2} \left[1 - \frac{d}{2}\left(\frac{1}{2 z}\right)^{d-2}\right]^{\frac{1}{d-1}} = b \,,
\end{eqnarray}
with $b$ the far away impact parameter. In $d = 3$, this leads to
\begin{eqnarray}
\fl
    f(z,w)\! = \! \Theta(-w)\! + \! \Theta(w)\!\left[\Theta\!\left(\!\sqrt{(1-w^2)z(4z-3)}\! - \! 1 \!\right)\! + \!\frac{1}{2} \delta\!\left(\!\sqrt{(1-w^2)z(4z-3)}\!-\!1\!\right)\!\right] .
\end{eqnarray}
Accordingly, the density can be obtained for $z>1$ as
\begin{eqnarray}
    \rho(z) = \frac{1}{2} + \frac{1}{2}\sqrt{\frac{z(4 z - 3)-1}{z(4 z - 3)}} + \frac{1}{4\sqrt{z(4 z - 3)}}\frac{1}{\sqrt{z(4z-3)-1}} \,.
\end{eqnarray}
This shows that the square-root divergence close to contact of the bulk density identified in Sec.~\ref{sec:vicinity} is robust to the inclusion of far-field hydrodynamic interactions, though it remains an open question to find out how near-field corrections would affect this behavior. The far-field structure of the density field is also left unchanged by hydrodynamic interactions as we obtain $\rho(z) \sim z^{-4}$ for $z\gg 1$, similarly to the $d=3$ case of Eq.~\eqref{eq:resultrho}. Lastly, the surface term can be deduced for $w<0$ from Eq.~\eqref{eq:surf8hydro} as
\begin{eqnarray}
    \Gamma(w) = \frac{1 - d 2^{1-d}}{d-1 - d 2^{1-d}} \,, \nonumber \\ \phantom{\Gamma(\theta)} = \frac{1}{d-1} + \frac{d 2^{1-d}(2-d)}{(d-1)(d-1-d 2^{1-d})} < \frac{1}{d-1} \,.
\end{eqnarray}
The amplitude of the sticking term is therefore reduced by far-field hydrodynamic interactions (from $1/2$ without hydrodynamic interactions downto $1/5$ when they are included, in $d=3$).

\section{Conclusion}\label{sec:conclusion}
We have studied analytically the steady-state of a run-and-tumble particle around a hard spherical obstacle. We have showed that as soon as the self-propulsion force has a non-vanishing correlation time, the RTP is effectively attracted to the obstacle when in its vicinity. This effect shows up in two ways. First, the steady-state distribution function displays a delta-peak accumulation at the surface of the obstacle. Second, the bulk density diverges with an exponent $-1/2$ close to it because the radial velocity of a particle leaving the obstacle vanishes, provided that it didn't flip its orientation in the course of skidding. We have then obtained the full steady-state distribution function in the limit where the persistence length of the RTP is much larger than the obstacle size. This also gives us the pair-distribution function at low density of a fluid of highly ballistic hard-core RTPs. We have furthermore showed that the amplitude of the delta-peak accumulation is an increasing function of the persistence length, at least when the latter is still large compared to the obstacle size. Lastly we have investigated the role of the far-field hydrodynamic interactions on colloids driven by an external force, as captured by the Oseen tensor.  We conjecture that hydrodynamic interactions reduce the propensity of highly persistent run-and-tumble particles to stick together while preserving the qualitative features of the bulk density. It would certainly be interesting to investigate the robustness of the structure of the distribution function not only upon incorporating near-field hydrodynamic effects but also by considering different propulsion mechanisms in the spirit of \cite{matas2014hydrodynamic}. Other stimulating open questions are concerned with non-spherical obstacles for which the far-field decay of the density field is known to be drastically different \cite{granek2020bodies}. The separation of the steady-state probability distribution into a bulk distribution and a surface distribution is expected to survive.\\

\bibliographystyle{unsrt}
\bibliography{biblio-activemf}

\begin{thebibliography}{10}

\bibitem{PhysRevLett.75.1226}
Tam\'as Vicsek, Andr\'as Czir\'ok, Eshel Ben-Jacob, Inon Cohen, and Ofer
  Shochet.
\newblock Novel type of phase transition in a system of self-driven particles.
\newblock {\em Phys. Rev. Lett.}, 75:1226--1229, Aug 1995.

\bibitem{gregoire2004onset}
Guillaume Gr{\'e}goire and Hugues Chat{\'e}.
\newblock Onset of collective and cohesive motion.
\newblock {\em Physical review letters}, 92(2):025702, 2004.

\bibitem{ramaswamy-2010}
Sriram Ramaswamy.
\newblock The mechanics and statistics of active matter.
\newblock {\em Annual Review of Condensed Matter Physics}, 1(1):323--345, 2010.

\bibitem{PhysRevLett.100.218103}
J.~Tailleur and M.~E. Cates.
\newblock Statistical mechanics of interacting run-and-tumble bacteria.
\newblock {\em Phys. Rev. Lett.}, 100:218103, May 2008.

\bibitem{redner2013structure}
Gabriel~S Redner, Michael~F Hagan, and Aparna Baskaran.
\newblock Structure and dynamics of a phase-separating active colloidal fluid.
\newblock {\em Physical review letters}, 110(5):055701, 2013.

\bibitem{PhysRevLett.108.235702}
Yaouen Fily and M.~Cristina Marchetti.
\newblock Athermal phase separation of self-propelled particles with no
  alignment.
\newblock {\em Phys. Rev. Lett.}, 108:235702, Jun 2012.

\bibitem{cates2015motility}
Michael~E Cates and Julien Tailleur.
\newblock Motility-induced phase separation.
\newblock {\em Annu. Rev. Condens. Matter Phys.}, 6(1):219--244, 2015.

\bibitem{Tailleur_2009}
J.~Tailleur and M.~E. Cates.
\newblock Sedimentation, trapping, and rectification of dilute bacteria.
\newblock {\em EPL (Europhysics Letters)}, 86(6):60002, Jun 2009.

\bibitem{pototsky2012active}
Andrey Pototsky and Holger Stark.
\newblock Active brownian particles in two-dimensional traps.
\newblock {\em EPL (Europhysics Letters)}, 98(5):50004, 2012.

\bibitem{solon2015active}
Alexandre~P Solon, Michael~E Cates, and Julien Tailleur.
\newblock Active brownian particles and run-and-tumble particles: A comparative
  study.
\newblock {\em The European Physical Journal Special Topics},
  224(7):1231--1262, 2015.

\bibitem{fodor2018non}
{\'E}tienne Fodor, Hisao Hayakawa, Julien Tailleur, and Fr{\'e}d{\'e}ric van
  Wijland.
\newblock Non-gaussian noise without memory in active matter.
\newblock {\em Physical Review E}, 98(6):062610, 2018.

\bibitem{ezhilan2015distribution}
Barath Ezhilan, Roberto Alonso-Matilla, and David Saintillan.
\newblock On the distribution and swim pressure of run-and-tumble particles in
  confinement.
\newblock {\em Journal of Fluid Mechanics}, 781, 2015.

\bibitem{wagner2017steady}
Caleb~G Wagner, Michael~F Hagan, and Aparna Baskaran.
\newblock Steady-state distributions of ideal active brownian particles under
  confinement and forcing.
\newblock {\em Journal of Statistical Mechanics: Theory and Experiment},
  2017(4):043203, 2017.

\bibitem{lee2013active}
Chiu~Fan Lee.
\newblock Active particles under confinement: aggregation at the wall and
  gradient formation inside a channel.
\newblock {\em New Journal of Physics}, 15(5):055007, 2013.

\bibitem{elgeti2015run}
Jens Elgeti and Gerhard Gompper.
\newblock Run-and-tumble dynamics of self-propelled particles in confinement.
\newblock {\em EPL (Europhysics Letters)}, 109(5):58003, 2015.

\bibitem{spagnolie2015geometric}
Saverio~E Spagnolie, Gregorio~R Moreno-Flores, Denis Bartolo, and Eric Lauga.
\newblock Geometric capture and escape of a microswimmer colliding with an
  obstacle.
\newblock {\em Soft Matter}, 11(17):3396--3411, 2015.

\bibitem{goldstein1950classical}
Herbert Goldstein, Charles Poole, and John Safko.
\newblock Classical mechanics. aw series in advanced physics, 1950.

\bibitem{van1984activation}
Christian Van~den Broeck and Peter H{\"a}nggi.
\newblock Activation rates for nonlinear stochastic flows driven by
  non-gaussian noise.
\newblock {\em Physical Review A}, 30(5):2730, 1984.

\bibitem{schnitzer1993theory}
Mark~J Schnitzer.
\newblock Theory of continuum random walks and application to chemotaxis.
\newblock {\em Physical Review E}, 48(4):2553, 1993.

\bibitem{solon2015Notpressure}
Alexandre~P Solon, Yaouen Fily, Aparna Baskaran, Mickael~E Cates, Yariv Kafri,
  Mehran Kardar, and J~Tailleur.
\newblock Pressure is not a state function for generic active fluids.
\newblock {\em Nature Physics}, 11(8):673--678, 2015.

\bibitem{dhar2019run}
Abhishek Dhar, Anupam Kundu, Satya~N Majumdar, Sanjib Sabhapandit, and
  Gr{\'e}gory Schehr.
\newblock Run-and-tumble particle in one-dimensional confining potentials:
  steady-state, relaxation, and first-passage properties.
\newblock {\em Physical Review E}, 99(3):032132, 2019.

\bibitem{thompson2011lattice}
Alasdair~G Thompson, Julien Tailleur, Michael~E Cates, and Richard~A Blythe.
\newblock Lattice models of nonequilibrium bacterial dynamics.
\newblock {\em Journal of Statistical Mechanics: Theory and Experiment},
  2011(02):P02029, 2011.

\bibitem{PhysRevLett.116.218101}
A.~B. Slowman, M.~R. Evans, and R.~A. Blythe.
\newblock Jamming and attraction of interacting run-and-tumble random walkers.
\newblock {\em Phys. Rev. Lett.}, 116:218101, May 2016.

\bibitem{malakar2018steady}
Kanaya Malakar, V~Jemseena, Anupam Kundu, K~Vijay Kumar, Sanjib Sabhapandit,
  Satya~N Majumdar, S~Redner, and Abhishek Dhar.
\newblock Steady state, relaxation and first-passage properties of a
  run-and-tumble particle in one-dimension.
\newblock {\em Journal of Statistical Mechanics: Theory and Experiment},
  2018(4):043215, 2018.

\bibitem{speck-kirkwood}
Andreas H\"artel, David Richard, and Thomas Speck.
\newblock Three-body correlations and conditional forces in suspensions of
  active hard disks.
\newblock {\em Phys. Rev. E}, 97:012606, Jan 2018.

\bibitem{Agoritsas21jsm}
Elisabeth Agoritsas.
\newblock Mean-field dynamics of infinite-dimensional particle systems: global
  shear versus random local forcing.
\newblock {\em Journal of Statistical Mechanics: Theory and Experiment},
  2021(3):033501, mar 2021.

\bibitem{Mo20sm}
Ruoyang Mo, Qinyi Liao, and Ning Xu.
\newblock Rheological similarities between dense self-propelled and sheared
  particulate systems.
\newblock {\em Soft Matter}, 16:3642--3648, 2020.

\bibitem{de2019active}
Thibaut~Arnoulx de~Pirey, Gustavo Lozano, and Fr{\'e}d{\'e}ric van Wijland.
\newblock Active hard spheres in infinitely many dimensions.
\newblock {\em Physical review letters}, 123(26):260602, 2019.

\bibitem{martin2021statistical}
David Martin, J{\'e}r{\'e}my O'Byrne, Michael~E Cates, {\'E}tienne Fodor,
  Cesare Nardini, Julien Tailleur, and Fr{\'e}d{\'e}ric van Wijland.
\newblock Statistical mechanics of active ornstein-uhlenbeck particles.
\newblock {\em Physical Review E}, 103(3):032607, 2021.

\bibitem{berner2018oscillating}
Johannes Berner, Boris M{\"u}ller, Juan~Ruben Gomez-Solano, Matthias
  Kr{\"u}ger, and Clemens Bechinger.
\newblock Oscillating modes of driven colloids in overdamped systems.
\newblock {\em Nature communications}, 9(1):999, 2018.

\bibitem{guazzelli2011physical}
Elisabeth Guazzelli and Jeffrey~F Morris.
\newblock {\em A physical introduction to suspension dynamics}, volume~45.
\newblock Cambridge University Press, 2011.

\bibitem{rotne1969variational}
Jens Rotne and Stephen Prager.
\newblock Variational treatment of hydrodynamic interaction in polymers.
\newblock {\em The Journal of Chemical Physics}, 50(11):4831--4837, 1969.

\bibitem{yamakawa1970transport}
Hiromi Yamakawa.
\newblock Transport properties of polymer chains in dilute solution:
  hydrodynamic interaction.
\newblock {\em The Journal of Chemical Physics}, 53(1):436--443, 1970.

\bibitem{felderhof1977hydrodynamic}
BU~Felderhof.
\newblock Hydrodynamic interaction between two spheres.
\newblock {\em Physica A: Statistical Mechanics and its Applications},
  89(2):373--384, 1977.

\bibitem{matas2014hydrodynamic}
Ricard Matas-Navarro, Ramin Golestanian, Tanniemola~B Liverpool, and Suzanne~M
  Fielding.
\newblock Hydrodynamic suppression of phase separation in active suspensions.
\newblock {\em Physical Review E}, 90(3):032304, 2014.

\bibitem{granek2020bodies}
Omer Granek, Yongjoo Baek, Yariv Kafri, and Alexandre~P Solon.
\newblock Bodies in an interacting active fluid: far-field influence of a
  single body and interaction between two bodies.
\newblock {\em Journal of Statistical Mechanics: Theory and Experiment},
  2020(6):063211, 2020.

\bibitem{brenner1981translational}
Howard Brenner.
\newblock The translational and rotational motions of an n-dimensional
  hypersphere through a viscous fluid at small reynolds numbers.
\newblock {\em Journal of Fluid Mechanics}, 111:197--215, 1981.

\end{thebibliography}

\appendix

\section{Dynamics in the $r,\,w$ variables}\label{app:dynvar}
The dynamics in Eq.~\eqref{eq:EOM} is closed in terms of the variables $r$ and $w$. We discuss the resulting dynamics for the case of a hard spherical obstacle. In between two tumbles, say at $t_0$ and $t_1$, and while the particle is away from the obstacle $r(t)\geq\sigma$, we have
\begin{equation}
r(t) = \sqrt{r(t_0)^2 + v_0^2\left(t-t_0\right)^2 + 2 v_0 \left(t-t_0\right)r(t_0)w(t_0)} \,,
\end{equation}
together with
\begin{equation}
w(t) = \frac{v_0 \left(t-t_0\right) + w(t_0)r(t_0)}{r(t)} \,.
\end{equation}
The last two equations are obtained by projecting Eq.~\eqref{eq:EOM} along $\bi{r}$ and $\bi{u}$. Let us now assume that the particle hits the obstacle at some intermediate time $t_2$ such that $r(t_2)=\sigma$. For all $t_2\leq t \leq t_1$ such that $w(t) < 0$, the particle skids on the obstacle and 
\begin{equation}
r(t) = \sigma \,,
\end{equation}
while
\begin{equation}
w(t) = \frac{e^{\frac{2v_0\left(t-t_0\right)}{\sigma}}(1+w(t_2)) + w(t_2) - 1}{e^{\frac{2v_0\left(t-t_0\right)}{\sigma}}(1+w(t_2)) + 1 -  w(t_2)} \,.
\end{equation}
Lastly, when a tumble occur, the new value of $w$ is sampled with the measure $P_{\rm tumble}(w) = \left(2 W_{d-2}\right)^{-1}\left(1-w^2\right)^{\frac{d-3}{2}}$. With increasing the dimension, $P_{\rm tumble}(w)$ gets more biased towards $w=0$, the counterpart of the increasing number of directions orthogonal to $\bi{r}$.

\section{Bound on the growth in equation \eqref{eq:equiv2}}\label{app:bound}
In this appendix, we show that if $\rho(z)$ is an integrable function at $z=1$, the integral in Eq.~\eqref{eq:equiv2}
\begin{equation}
I = \int^a_h \rmd h' \rho(1+h')\ln(h') \,,
\end{equation}
does not grow faster than $\ln h$ as $h\to 0$ for any finite $a>0$. Let 
\begin{equation}
P(h) = \int_0^h \rho(1+h') \rmd h' \,.
\end{equation}
It is such that $P(h)\to 0$ as $h\to 0$. By integration by part we have,
\begin{equation}
I = P(a)\ln a - P(h)\ln h - \int_h^a \frac{P(h')}{h'} \,.
\end{equation}
We can lastly write
\begin{equation}
\int_h^a \frac{P(h')}{h'} < P(a)\left(\ln a - \ln h\right)\,,
\end{equation}
that proves that $I$ does not grow faster than $\ln h$ as $h\to 0$.
\section{Derivation of the equations of motion with hydrodynamic interactions}\label{app:Oseen}
\subsection{Fluid flow around a moving obstacle}\label{subsec:hydro1}
We consider a moving spherical obstacle of radius $\sigma/2$ at velocity $\bi{u}$. The fluid is incompressible and viscous with viscosity $\mu$ and the flow is given by the Stokes equation with no-slip boundary conditions
\begin{equation}
    \begin{cases}{
        \mu \Delta \bi{v} - \bnabla P = 0 \,, \\ 
        \bnabla \cdot \bi{v} = 0 \,, \\
        \bi{v}(\bi{r} = (\sigma/2) \hat{\bi{r}}) = \bi{u} \,, \\
        P(\bi{r}) = P_{\infty} \,\, {\rm as} \,\, ||\bi{r}|| \to \infty \,. \\}
    \end{cases}
\end{equation}
We hereafter follow \cite{brenner1981translational}. Due to the linearity of the equations in $\bi{u}$, we introduce $M^{\alpha}_\beta$ and $X_\beta$ such that
\begin{equation}
    \begin{cases}{
         v^\alpha(\bi{r}) = M^\alpha_\beta(\bi{r}) \bi{u}^\beta \,, \\
         P(\bi{r}) - P_\infty = \eta X_\beta(\bi{r}) \bi{u}^\beta \,. \\}
    \end{cases}
\end{equation}
They satisfy
\begin{equation}
    \begin{cases}{
         \partial_\beta\partial^\beta M^\alpha_\gamma - \partial^\alpha X_\gamma = 0  \,,\\
         \partial_\alpha M^\alpha_\gamma = 0 \,,\\
         M^\beta_\gamma (\bi{r} = (\sigma/2) \hat{\bi{r}}) = \delta^\beta_\gamma \,, \\
         X_\beta(\bi{r}) = 0 \,\, {\rm as} \,\, ||\bi{r}|| \to \infty \,, \\}
    \end{cases}
    \label{brenner}
\end{equation}
which is a completely rotationaly invariant set of equations. We have hence by symmetry
\begin{equation}
    \begin{cases}{
        X_\beta(\bi{r}) = f(r) \bi{r}_\beta \,, \\
        M^\alpha_\gamma(\bi{r}) = M_1(r) \delta^\alpha_\gamma + M_2(r) \bi{r}^\alpha \bi{r}_\gamma \,. \\}
    \end{cases}
\end{equation}
We can therefore obtain a set of three coupled equations for $f$, $M_1$ and $M_2$ as
\begin{equation}
    \begin{cases}{
       M_1'' + \frac{d-1}{r}M_1' + 2M_2 - f = 0 \,\, , \, M_1(\sigma/2) = 1 \,, \\
       M_2'' + \frac{d-1}{r}M_2' + \frac{4 M_2'}{r} - \frac{f'}{r} = 0 \,\, , \, M_2(\sigma/2) = 0 \,, \\
       \frac{M_1'}{r} + r M_2' + (d+1)M_2 = 0 \,. \\}
    \end{cases}
\label{eq:MF}
\end{equation}
Combining them so as to express the vanishing of the Laplacian of the pressure field, we arrive at 
\begin{equation}
    f'' + \frac{d+1}{r}f' = 0 \,,
\end{equation}
so that
\begin{equation}
    f(r) = \frac{A}{r^d} \,.
\end{equation}
When injecting this expression into the second equation of Eq.~\eqref{eq:MF}, we obtain $M_2(r)$ as
\begin{equation}
    M_2(r) = \frac{A}{2 r^d}\left(1- \frac{\sigma^2}{4 r^2}\right) \,,
\end{equation}
which in turn can eventually be plugged into the first equation of Eq.~\eqref{eq:MF} so as to get
\begin{equation}
    M_1(r) = \frac{A}{2d}\left(\frac{\sigma^2}{4 r^d}- \frac{1}{r^{d-2}}\right) + \frac{\sigma^{d-2}}{2^{d-2}r^{d-2}} \,.
\end{equation}
The constant $A$ can then be obtained thanks to the divergenceless condition of the velocity field. This eventually yields the pressure and velocity fields which are given by
\begin{equation}
\fl
	\begin{cases}{
     P(\bi{r}) = P_\infty + \mu \frac{d(d-2)}{d-1}\frac{2}{\sigma}\left(\frac{\sigma}{2 r}\right)^{d-1}\hat{\bi{r}}\cdot \bi{u} \,, \\
    \bi{v}(\bi{r}) = \frac{1}{2(d-1)}\left(\frac{\sigma}{2 r}\right)^d \left[d \left(\frac{2 r}{\sigma}\right)^2 + (d-2)\right]\bi{u} + \frac{d(d-2)}{2(d-1)}\left(\frac{\sigma}{2 r}\right)^d \left[\left(\frac{2 r}{\sigma}\right)^2 -1\right]\left(\hat{\bi{r}}\cdot \bi{u}\right)\hat{\bi{r}} \,. \\} 
    \end{cases}
\end{equation}
\subsection{Two moving spheres and hydrodynamic interactions}
We now consider the case of two spheres of radius $\sigma/2$ at positions $\bi{r}_1$ and $\bi{r}_2$ with velocities $\bi{u}_1$ and $\bi{u}_2$. The flow field is given by the solution of the Stokes equation subjected to the proper boundary conditions
\begin{equation}
    \begin{cases}{
       \mu \Delta \bi{v} - \bnabla P = 0 \,, \\
       \bnabla \cdot \bi{v} = 0 \,, \\
       \bi{v}(\bi{r}_1 + (\sigma/2)\hat{\bi{r}}) = \bi{u}_1 \,, \\ 
       \bi{v}(\bi{r}_2 + (\sigma/2)\hat{\bi{r}}) = \bi{u}_2 \,. \\}
    \end{cases}
\end{equation}
In this case, we look for the solution as
\begin{equation}
\begin{cases}{
    \bi{v}(\bi{r}) = \bi{v}_1^{(0)}(\bi{r}-\bi{r}_1) + \bi{v}_2^{(0)}(\bi{r}-\bi{r}_2) + \bi{v}_{corr}(\bi{r}) \,, \\ 
    P(\bi{r}) = P_0^{(0)}(\bi{r}-\bi{r}_1) +  P_0^{(0)}(\bi{r}-\bi{r}_2) + P_{corr}(\bi{r}) + P_\infty \,, \\}
\end{cases}
\end{equation}
with $\bi{v}_1^{(0)}(\bi{r}-\bi{r}_1)$ the fluid flow due to particle 1 in absence of particle 2 (and accordingly for particle 2 and for the pressure fields). The correcting terms satisfy the following system of equations
\begin{equation}
    \begin{cases}{
       \mu \Delta \bi{v}_{corr} - \bnabla P_{corr} = 0 \,, \\
       \bnabla \cdot \bi{v}_{corr} = 0 \,, \\
       \bi{v}_{corr}(\bi{r}_1 + (\sigma/2)\hat{\bi{r}}) = - \bi{v}_2^{(0)}(\bi{r}_1-\bi{r}_2 + (\sigma/2)\hat{\bi{r}}) \,,  \\ 
       \bi{v}_{corr}(\bi{r}_2 + (\sigma/2)\hat{\bi{r}}) = - \bi{v}_1^{(0)}(\bi{r}_2-\bi{r}_1 + (\sigma/2)\hat{\bi{r}}) \,. \\}
    \end{cases}
\end{equation}
The correcting velocity field $\bi{v}_{corr}$ vanishes as the separation between the two particles goes to infinity (comparatively to their size). In the following we will be interested in the first correction of $v_{corr}$ in the inverse separation $\sigma/||\bi{r}_1-\bi{r}_2||$. We define $ \delta \bi{v}_2 = \bi{v}_2^{(0)}(\bi{r}_1-\bi{r}_2 + \sigma/2\hat{\bi{r}})$, and accordingly $\delta \bi{v}_1 = \bi{v}_1^{(0)}(\bi{r}_2-\bi{r}_1 + \sigma/2\hat{\bi{r}})$, which, from \ref{subsec:hydro1},  writes to leading order in the inverse separation,
\begin{eqnarray}
\fl       \delta \bi{v}_2 =  \frac{d(d-2)}{2(d-1)}\left(\frac{\sigma}{2|\bi{r}_1-\bi{r}_2|}\right)^{d-2} \frac{\left((\bi{r}_1-\bi{r}_2)\cdot\bi{u}_2\right)(\bi{r}_1-\bi{r}_2)}{|\bi{r}_1-\bi{r}_2|^2} \\ \fl \phantom{\delta \bi{v}_2 = }  + \frac{d}{2(d-1)}\left(\frac{\sigma}{2|\bi{r}_1-\bi{r}_2|}\right)^{d-2} \bi{u}_2  + O\left(\left(\frac{\sigma}{2|\bi{r}_1-\bi{r}_2|}\right)^{d-1}\right)  \,,
\end{eqnarray}
which is therefore independent of $\hat{\bi{r}}$. As a consequence, up to their first order correction in the inverse separation, the velocity and pressure fields are given by the sum of the fields sourced by two isolated spherical particles with shifted velocities: one at $\bi{r}_1$ with velocity $\bi{u}_1 - \delta \bi{v}_2$ and one at $\bi{r}_2$ with velocity $\bi{u}_2 - \delta \bi{v}_1$. Hence 
\begin{eqnarray}
\fl 
    \bi{v}(\bi{r}) =  \frac{1}{2(d-1)}\left(\frac{\sigma}{2 |\bi{r} -\bi{r}_1|}\right)^d \left[d \left(\frac{2 |\bi{r}_1 - \bi{r}|}{\sigma}\right)^2 + (d-2)\right](\bi{u}_1-\delta \bi{v}_2) \nonumber \\ \fl \phantom{\bi{v}(\bi{r}) = } + \frac{1}{2(d-1)}\left(\frac{\sigma}{2 |\bi{r} -\bi{r}_2|}\right)^d \left[d \left(\frac{2 |\bi{r}_2 - \bi{r}|}{\sigma}\right)^2 + (d-2)\right](\bi{u}_2-\delta \bi{v}_1)  \nonumber \\ \fl \phantom{\bi{v}(\bi{r}) = } + \frac{d(d-2)}{2(d-1)}\left(\frac{\sigma}{2 |\bi{r}_1 - \bi{r}|}\right)^d \left[\left(\frac{2 |\bi{r}_1 - \bi{r}|}{\sigma}\right)^2 -1\right]\frac{\left((\bi{r}-\bi{r}_1)\cdot (\bi{u}_1 - \delta \bi{v}_2)\right)(\bi{r}-\bi{r}_1)}{|\bi{r}-\bi{r}_1|^2}  \nonumber \\ \fl \phantom{\bi{v}(\bi{r}) = } + \frac{d(d-2)}{2(d-1)}\left(\frac{\sigma}{2 |\bi{r}_2 - \bi{r}|}\right)^d \left[\left(\frac{2 |\bi{r}_2 - \bi{r}|}{\sigma}\right)^2 -1\right]\frac{\left((\bi{r}-\bi{r}_2)\cdot (\bi{u}_2 - \delta \bi{v}_1)\right)(\bi{r}-\bi{r}_2)}{|\bi{r}-\bi{r}_2|^2} \,.
\end{eqnarray}
Accordingly, the pressure field is given by
\begin{eqnarray}
  \fl  P(\bi{r}) = \, P_\infty + \, \mu \frac{d(d-2)}{d-1}\frac{2}{\sigma}\left(\frac{\sigma}{2 |\bi{r}-\bi{r}_1|}\right)^{d-1}\frac{(\bi{r}-\bi{r}_1)\cdot(\bi{u}_1 - \delta \bi{v}_2)}{|\bi{r}-\bi{r}_1|} \nonumber \\  \fl  \phantom{P(\bi{r}) =} + \mu \frac{d(d-2)}{d-1}\frac{2}{\sigma}\left(\frac{\sigma}{2 |\bi{r}-\bi{r}_2|}\right)^{d-1}\frac{(\bi{r}-\bi{r}_2)\cdot(\bi{u}_2 - \delta \bi{v}_1)}{|\bi{r}-\bi{r}_2|} \,.
\end{eqnarray}
We now seek to compute the force exerted by the fluid flow on particle 1,
\begin{eqnarray}
    \bi{F} = \int \rmd \mathbb{S} \cdot \mathbb{\pi} \,,
\end{eqnarray}
with $\pi_{ij} = - P \delta_{ij} + \eta \left(\partial_i v_j + \partial_j v_i\right)$ the stress tensor. One can check that to next to leading order in the relative separation, the force exerted by the fluid on particle 1 is the same as the one exerted on an isolated particle with velocity $\bi{u}_1 - \delta \bi{v}_2$, \textit{i.e.}
\begin{eqnarray}
    \bi{F} = - \mu \frac{ d(d-2)}{d-1} \frac{\Omega_d\sigma^{d-2}}{2^{d-2}}(\bi{u}_1 - \delta \bi{v}_2) \,.
\end{eqnarray}
The equations of motion \eqref{eq:EOM_hydro} of the main text then follow.

\end{document}